\documentclass[%
groupedaddress,
 aip,
jcp,%
 amsmath,amssymb,
 reprint,%
]{revtex4-1}

\usepackage{graphicx}%
\usepackage{dcolumn}%
\usepackage{bm}%
\usepackage{xcolor}
\usepackage{siunitx}

\newcommand{\Tr}{\mathrm{Tr}}
\newcommand{\tr}{\mathrm{tr}}
\newcommand{\rd}{\mathrm{d}}
\newcommand{\ii}{\mathrm{i}}

\newcommand{\eu}[1]{\mathrm{e}^{#1}}
\newcommand{\pd}[2]{\frac{\partial #1}{\partial #2}}
\newcommand{\thalf}{\ensuremath{\tfrac{1}{2}}}
\renewcommand{\Re}{\mathrm{Re}}
\renewcommand{\Im}{\mathrm{Im}}
\providecommand{\mat}[1]{{#1}}
\usepackage{soul} %
\newcommand{\Ref}[1]{Ref.~\onlinecite{#1}}
\newcommand{\Refs}[1]{Refs.~\onlinecite{#1}}
\newcommand{\eqn}[1]{Eq.~\eqref{eq:#1}}

\usepackage{braket}

\usepackage{dcolumn} %
\newcolumntype{d}[1]{D{.}{.}{#1}} %

\begin{document}

\title{Generalized spin mapping for quantum-classical dynamics}%

\author{Johan E. Runeson}
\email{johan.runeson@phys.chem.ethz.ch}
\author{Jeremy O. Richardson}%
\email{jeremy.richardson@phys.chem.ethz.ch}
\affiliation{Laboratory of Physical Chemistry, ETH Z\"{u}rich, 8093 Z\"{u}rich, Switzerland}

\date{\today}%

\begin{abstract}
We recently derived a spin-mapping approach for treating the nonadiabatic dynamics of a two-level system in a classical environment [\emph{J. Chem. Phys.} \textbf{151}, 044119 (2019)] based on the well-known quantum equivalence between a two-level system and a spin-1/2 particle.
In the present paper, we generalize this method
to describe the dynamics of $N$-level systems.
This is done via a mapping to a classical phase space that preserves the $SU(N)$-symmetry of the original quantum problem.
The theory reproduces the standard Meyer--Miller--Stock--Thoss Hamiltonian without invoking an extended phase space, and we thus avoid leakage from the physical subspace. In contrast with the standard derivation of this Hamiltonian, the generalized spin mapping leads to an $N$-dependent value of the zero-point energy parameter that is uniquely determined by the Casimir invariant of the $N$-level system. 
Based on this mapping, we derive a simple way to approximate correlation functions in complex nonadiabatic molecular systems via classical trajectories, and present benchmark calculations on the seven-state Fenna--Matthews--Olson complex. The results are significantly more accurate than conventional Ehrenfest dynamics, at a comparable computational cost, and can compete in accuracy with other state-of-the-art mapping approaches.

\end{abstract}

\maketitle

\section{\label{sec:intro}Introduction}
The full quantum dynamics of complex systems is in general far too complicated to be simulated in practice. Instead it is often necessary to separate the problem into a (smaller) subsystem that is treated  quantum-mechanically and an environment that can be approximated by classical dynamics. In chemistry, the typical example is to treat a molecular system as a subsystem of $N$ electronic levels coupled to an environment of classical nuclear modes. If the coupling between the electronic and nuclear motion cannot be neglected, methods based on the standard Born--Oppenheimer approximation are not applicable. Instead new methods are needed to describe such \emph{nonadiabatic} processes, which are important for the study of solar cells, vision, and photosynthesis, among others.\cite{Tully2012perspective} 

One way to make large-scale simulations of these phenomena possible is to approximate the nuclear motion by an ensemble of independent trajectories that propagate under classical equations of motion.
Among the simplest trajectory-based methods are Ehrenfest dynamics, in which the nuclei move on a mean-field potential defined by the instantaneous electronic populations, while the electronic variables follow exact subsystem dynamics according to the instantaneous nuclear configuration. This method has a number of known severe drawbacks,\cite{grunwaldQCLEchapter} but is still popular due to its simplicity and low computational cost. Other options of comparable cost include surface hopping\cite{Tully1990hopping} and mapping-based techniques. \cite{Stock2005nonadiabatic}
In particular, the Meyer--Miller--Stock--Thoss (MMST) mapping\cite{Meyer1979nonadiabatic,Stock1997mapping} has recently regained attention.\cite{Cotton2013a,Miller2016Faraday,liu2016,geva2018,saller2019jcp} 
As a generalization of the Schwinger bosonization to $N$-level systems, its basic principle is to represent the $N$ electronic states by $N$ coupled harmonic oscillators that share a single excitation. This mapping is formally exact and has inspired a number of methods for calculating correlation functions, such as the linearized semiclassical initial-value representation (LSC-IVR),\cite{sun1998semiclassical} the Poisson-bracket mapping equation (PBME),\cite{Kim2008Liouville,Kelly2012mapping} the symmetrical quasiclassical windowing approach (SQC),\cite{Cotton2013a,Cotton2013b} partially linearized density matrix dynamics (PLDM),\cite{Huo2011densitymatrix,Huo2013PLDM} and the forward-backward trajectory solution (FBTS)\cite{Hsieh2012FBTS,Hsieh2013FBTS} of the quantum-classical Liouville equation.\cite{kapral1999mixed}
These \emph{quasiclassical} approaches all use a classical description of the nuclear dynamics, while preserving the exact quantum dynamics of an isolated subsystem.

Even though the MMST mapping is formally exact, its descendant methods are not, due to the quasiclassical approximation. 
In particular the classical dynamics may bring the system out of the singly-excited subspace.\cite{Kelly2012mapping} One way to improve upon this is to introduce additional projectors. In principle one could do this at every time step, but in practice this is usually done only at the start and/or end of the simulation.\cite{hsieh2013analysis} Another problem is that the zero-point energy of the fictitious harmonic oscillators is not respected by the classical dynamics. Historically it has been observed that this leakage can be mitigated by reducing the zero-point energy from 1 to a parametric value $\gamma$.\cite{muller1999a,Mueller1999pyrazine} In the more recently introduced symmetrical quasiclassical windowing approach (SQC),\cite{Cotton2013a} $\gamma$ is determined via a window function, which is in turn freely chosen. In the case of two-level systems, there is a natural choice of $\gamma$ that originates from the mapping of a spin vector, which was first proposed by Cotton and Miller\cite{Cotton2013b} and was derived in our previous paper (paper I)\cite{runeson2019} by mapping the two-level system to a spin-$\thalf$ instead of two harmonic oscillators. In the present paper we show that this spin mapping can be generalized to multiple levels. Its dynamics turns out to be equivalent to that of the MMST Hamiltonian, but with a new zero-point energy parameter $\gamma$, for which we derive a closed formula as a function of the number of levels.

The search for such a theory follows the intuition of Meyer and Miller, who originally considered the well-known equivalence between a two-level system and a spin-$\thalf$ system as an alternative derivation of their method.\cite{meyer1979spin} This however turned out to be difficult to extend to many levels. Since their generalization no longer reduced to give the correct dynamics for an isolated subsystem, they abandoned this path in favour of the harmonic-oscillator mapping, which since then has inspired the rich field of mapping-based methods mentioned above.
More recently, Cotton and Miller returned to the idea of a spin mapping by representing the two-level problem in terms of \emph{two} spins, in the hope of finding a more natural mapping than to harmonic oscillators.\cite{Cotton2015}  %
Unfortunately, this approach did not reduce to the correct dynamics for isolated subsystems either, which has lead some authors to believe that spin is not a good classical analogue for a quantum system.\cite{liu2016} In the present paper we demonstrate how a spin mapping can indeed be generalized to multi-level systems, in a way that
gives identical results to the Schr\"{o}dinger equation for an isolated subsystem.

The main practical difference between our spin mapping and the MMST mapping lies in the definition of the phase-space distribution. While the $2N$-dimensional phase space of MMST is unbounded, the spin-mapping phase space is confined to a sphere with $2N-2$ degrees of freedom. In this way, this phase space conserves the symmetries of the original quantum problem. The phase-space construction used in spin mapping was originally proposed by Stratonovich,\cite{stratonovich1957distributions} and is now known as the Stratonovich--Weyl (SW) representation, which has found various applications in quantum optics.\cite{klimov2009group} It is a generalization of Weyl's correspondence rule\cite{Weyl1927} and the classical phase-space theories by Wigner and Moyal.\cite{wigner1932,moyal1949quantum} Early works of SW-representations for spin were made by Agarwal, V{\'a}rilly and Gracia-Bond{\'\i}a.\cite{agarwal1981,varilly1989moyal} These rely on the properties of the $SU(2)$ Lie group, the fundamental symmetry of particles with spin. Brif and Mann have presented a construction for general Lie groups\cite{brif1999phase} and later Klimov and de Guise\cite{Klimov2010} as well as Tilma and Nemoto\cite{Tilma2012} for the case of $SU(N)$, which is the symmetry group of $N$-level systems.
This has recently been used in the study of qudits (qubits generalized to multiple states).\cite{Tilma2016,rundle2017,rundle2019, marchiolli2019}

In this paper we apply the Stratonovich--Weyl formalism to describe nonadiabatic dynamics in $N$-level molecular systems (but the resulting method is applicable for any quantum-classical problem). This leads to a straight-forward generalization of our results for the two-level system.\cite{runeson2019} The Stratonovich--Weyl representations could be formulated in spherical variables of coherent states, but like in the two-level case there is also a natural description in Cartesian variables, which leads to the same form of the Hamiltonian as in the MMST mapping, but with a more natural phase space that does not require projections and cannot suffer from unphysical leakage. In particular we derive a previously unknown closed formula for the zero-point energy parameter $\gamma$ in terms of $N$.%

The generalized spin mapping is not just a useful methodology in itself, but may also give insights about the standard MMST mapping. Recently it was found that the accuracy of MMST-based methods like LSC-IVR and PBME can be significantly improved by separating all observables into a linear combination of the identity operator and a traceless operator.\cite{saller2019jcp,saller2020faraday} While it is not so obvious from the harmonic-oscillator picture why this would be a more natural choice, it is clear from the construction of the spin mapping that the identity must be treated separately. Therefore, the key to understand the success of traceless MMST might lie in the generalized spin mapping.

In Sec.~\ref{sec:theory} we use the generalized spin mapping to approximate correlation functions in a manner similar to classical Wigner dynamics. %
In Sec.~\ref{sec:FMO} we apply the method to the seven-state Fenna--Matthews--Olson complex, which is a benchmark problem relevant for studies of light harvesting. The results can compete in accuracy with other state-of-the-art methods in the mapping community, and are far superior to conventional Ehrenfest dynamics with comparable cost.

\section{Theory}\label{sec:theory}
Consider a molecular system with $N$ electronic states and the general diabatic Hamiltonian
\begin{align} \label{eq:hamiltonian}
    \hat{H} &= \frac{\hat{p}^2}{2m} + \hat{V}(\hat{x}),
\end{align}
where $\hat{x}$ and $\hat{p}$ are vectors of position and momentum operators of the nuclear modes with associated mass $m$, and $\hat{V}(x)$ is a Hermitian potential-energy matrix of shape $N\times N$. We use the diabatic representation in this paper since it leads to the simplest formulation, but working in the adiabatic representation would also be possible.\cite{cotton2017adiabatic}

Like in other trajectory-based methods, we will treat the nuclear variables classically (that is, replace $\hat{x},\hat{p}\mapsto x,p$) but keep the quantum-mechanical evolution of the electronic operators. To handle the coupling between the two in a consistent fashion, we will map the electronic (subsystem) operators to a phase-space representation in which all variables are treated on the same footing. The mapping procedure will be similar to the spin mapping for two levels in paper I.\cite{runeson2019} In each section we will therefore first remind the reader of the two-level case, before generalizing to $N$ levels.

Throughout this paper we set $\hbar=1$.

\subsection{Generalization of the spin matrices}\label{sec:spinmatrices}
First we discuss the spin matrix decomposition of two-level Hamiltonians, before we generalize to $N$ levels. Consider the Hamiltonian in Eq.~\eqref{eq:hamiltonian} with a general (diabatic) potential matrix:
\begin{equation}\label{eq:Vmatrix}
   \hat{V}(x)   =  \begin{pmatrix} V_1(x) & \Delta^*(x) \\ \Delta(x) & V_2(x) \end{pmatrix}.
\end{equation}
It is well known that the Hamiltonian, or any other two-level Hermitian operator, can be decomposed into a basis of spin operators and the identity:
\begin{align} \label{eq:Hspinrepr}
    \hat{H} &= H_0 \hat{\mathcal{I}} + H_1 \hat{S}_1 + H_2 \hat{S}_2 + H_3\hat{S}_3 \\
    &= H_0 \hat{\mathcal{I}} + \bm{H}\cdot \hat{\bm{S}}, \nonumber
\end{align}
where %
\begin{equation*} \label{eq:spinops}
    \hat{S}_1 = \frac{1}{2}\begin{pmatrix} 0 & 1 \\ 1 & 0 \end{pmatrix},~ \hat{S}_2 =\frac{1}{2} \begin{pmatrix} 0 & -\ii \\ \ii & 0 \end{pmatrix},~ \hat{S}_3 =\frac{1}{2} \begin{pmatrix} 1 & 0 \\ 0 & -1 \end{pmatrix},
\end{equation*}
are the Pauli matrices multiplied by $\thalf$. The explicit relations between the quantities in Eqs.~\eqref{eq:hamiltonian} and \eqref{eq:Hspinrepr} are:
\begin{subequations}
\begin{align}
    H_0 &= \frac{p^2}{2m} + \tfrac{1}{2}(V_1(x)+V_2(x)) \\
    H_1 &= 2 \,\mathrm{Re} \,\Delta(x) \\
    H_2 &= 2 \,\mathrm{Im} \,\Delta(x) \\
    H_3 &= V_1(x)-V_2(x).
\end{align}
\end{subequations}
Without loss of generality we choose $\hat{V}(x)$ to be real, so that $H_2=0$.

Let us point out three important properties of the spin operators. 
First, they are traceless (i.e.\ $\tr[\hat{S}_i]=0$) in contrast to $\hat{\mathcal{I}}$ that has $\tr[\hat{\mathcal{I}}]=2$, where lowercase $\tr$ denotes a trace over the subsystem degrees of freedom).
As a consequence, the trace of $\hat{V}(x)$ will only appear in $H_0$, while $H_{i\geq 1}$ only depends on the traceless part of $\hat{V}(x)$. 
Note that this appears naturally and is not artificially imposed on the mapping, as is sometimes necessary for other mappings.\cite{Cotton2013b,Kelly2012mapping,saller2019jcp} %

Second, the spin matrices are orthogonal:
\begin{equation}\label{eq:trSiSj}
    \tr[\hat{S}_i\hat{S}_j] = \frac{1}{2} \delta_{ij}.
\end{equation}
Other normalizations of the spin matrices are possible, but we shall keep the factor of 1/2 to maintain the connection to a spin system.

The third relevant property of the spin matrices is that the sum of their squares is proportional to the identity:
\begin{equation}
    \sum_{i=1}^3 \hat{S}_i^2 = \frac{3}{4}\hat{\mathcal{I}}.
\end{equation}
The reader probably recognizes the square-root of the proportionality constant, $\sqrt{3}/2=\sqrt{S(S+1)}$, as the magnitude of a classical spin vector for a spin $S=1/2$. This observation will be important in the treatment of the $N$-level system.

Let us now generalize to an $N$-level potential. A general Hermitian $(N\times N)$-matrix has $N^2$ independent elements, or $N^2-1$ for traceless matrices. Therefore the basis expansion can be written on the form
\begin{equation}\label{eq:Sbasis}
    \hat{H} = H_0 \hat{\mathcal{I}} + \sum_{i=1}^{N^2-1} H_i \hat{S}_i,
\end{equation}
where $\hat{S}_i$ are now $(N\times N)$-matrices (also called the \emph{generators} of the $\mathfrak{su}(N)$ Lie algebra). The matrices $\hat{S}_i$ are necessarily traceless, and we keep the same normalization as in the two-level case, such that Eq.~\eqref{eq:trSiSj} is still fulfilled. %
Finally, it is well-known in the literature that the sum of the squares of the basis matrices is
\begin{equation}\label{eq:Casimir}
    \sum_{i=1}^{N^2-1} \hat{S}_i^2 = \frac{N^2-1}{2N}\hat{\mathcal{I}},
\end{equation}
which is called the (quadratic) \emph{Casimir operator} of $\mathfrak{su}(N)$. We include a short proof in Appendix~\ref{sec:app_casimir} for completeness. The Casimir operator is invariant to unitary basis transformations, and therefore not dependent on the particular choice of decomposition in Eq.~\eqref{eq:Sbasis}.
This simple expression will be the key to defining the zero-point energy parameter, which ultimately leads to significant improvements upon the MMST-mapping results.

There are many possible ways to choose the basis matrices,\cite{alicki2007semigroups,bertlmann2008qudits} but the theory of this paper will not depend on this choice.
As an example for $N=3$, a direct generalization of the Pauli matrices are the Gell-Mann matrices (which have been used in the $SU(3)$-symmetric theory of quarks\cite{gellmann1962}):
\begin{subequations}\label{eq:GellMann}
\begin{align*}
\hat{S}_1 &= \frac{1}{2}\begin{pmatrix}
0 & 1 & 0 \\ 1 & 0 & 0 \\ 0 & 0 & 0
\end{pmatrix}~ \hat{S}_2 = \frac{1}{2}\begin{pmatrix}
0 & -\ii & 0 \\ \ii & 0 & 0 \\ 0 & 0 & 0
\end{pmatrix}~ \hat{S}_3 = \frac{1}{2}\begin{pmatrix}
1 & 0 & 0 \\ 0 & -1 & 0 \\ 0 & 0 & 0
\end{pmatrix} \\
\hat{S}_4 &= \frac{1}{2}\begin{pmatrix}
0 & 0 & 1 \\ 0 & 0 & 0 \\ 1 & 0 & 0
\end{pmatrix} ~ \hat{S}_5 = \frac{1}{2}\begin{pmatrix}
0 & 0 & -\ii \\ 0 & 0 & 0 \\ \ii & 0 & 0
\end{pmatrix} ~
\hat{S}_6 = \frac{1}{2}\begin{pmatrix}
0 & 0 & 0 \\ 0 & 0 & 1 \\ 0 & 1 & 0
\end{pmatrix} \\
\hat{S}_7 &= \frac{1}{2}\begin{pmatrix}
0 & 0 & 0 \\ 0 & 0 & -\ii \\ 0 & \ii & 0
\end{pmatrix} \quad
\hat{S}_8 = \frac{1}{2\sqrt{3}}\begin{pmatrix}
1 & 0 & 0 \\ 0 & 1 & 0 \\ 0 & 0 & -2
\end{pmatrix},
\end{align*}
\end{subequations}
and the reader can easily confirm that $\sum_i \hat{S}_i^2=\frac{3^2-1}{2\cdot 3}\hat{\mathcal{I}}=\frac{4}{3}\hat{\mathcal{I}}$.
Note that the first three contain the two-level basis matrices as blocks padded with zeros.
This construction can be generalized to higher $N$, and the details are given in Appendix~\ref{sec:app_basis}. In practice, we shall see that it is not necessary for the results of this paper to carry out the expansion in Eq.~\eqref{eq:Sbasis} at all, so that the basis does not have to be known explicitly.

One might ask what the basis matrices have to do with spins once $N>2$. Already in the 1970s, Meyer and Miller proposed a mapping of the $N$-level system to a higher spin $S=\frac{1}{2}(N-1)$.\cite{meyer1979spin} While their spin-matrix decomposition is the same as ours for $N=2$, it is different for all $N>2$, since in their construction not all $\hat{S}_i$ are traceless (such that there is no Casimir invariant as in Eq.~\eqref{eq:Casimir}). Their basis matrices are therefore not generators of $\mathfrak{su}(N)$ and cannot be used to derive the results of this paper.
Nonetheless, this can be easily fixed such that it is possible to construct the $\mathfrak{su}(N)$ generators from a spin picture. For the interested reader we show in Appendix~\ref{sec:app_spin1} how to obtain basis matrices for $N=3$ by describing a spin-1 system as two interacting spin-$\thalf$ particles in a triplet configuration.

An even more important difference between the approach introduced in this paper and that of \Ref{meyer1979spin} is the phase-space representation used to convert the spin matrices to classical variables.
Meyer and Miller mapped each matrix to the same two variables for any number of levels, which again is equivalent to our work for the $N=2$ case,\footnote{In their notation, $q=\varphi$ and $m=\thalf\cos\theta$.} but not for $N>2$.
Although a two-variable phase-space is appropriate for the $SU(2)$ symmetry of a single spin-1 particle,
the true symmetry group of the three-level system is $SU(3)$, which we represent by four phase-space variables, as explained in Sec.~\ref{sec:SW}.

Thoss and Stock have also investigated a spin-$\thalf$ mapping of two-level systems,\cite{Thoss1999mapping} and derived a semiclassical initial-value representation of its corresponding propagator. Like our approach, they also use spin coherent states, and their dynamics is exact for an isolated subsystem. However, they did not generalize their method to more than two levels. In this paper we pursue a quasiclassical approach to such a generalization by using the Stratonovich--Weyl representation of the $N$-level problem.

\subsection{Stratonovich--Weyl representations}\label{sec:SW}
Again, we will start with the two-level case that was previously presented in paper I.\cite{runeson2019} As is commonly done in textbooks, one can think of the diabatic states $|1\rangle$ and $|2\rangle$ as the eigenstates of a (fictitious) spin-$\thalf$ degree of freedom.
In order to map these to a phase-space, we introduce the spin-$\thalf$ coherent states\cite{Radcliffe1971}
\begin{equation}\label{eq:coherent_state2}
    |\bm{u}\rangle = \cos\tfrac{\theta}{2}\,\eu{-\ii\varphi/2}|1\rangle + \sin\tfrac{\theta}{2}\,\eu{\ii\varphi/2}|2\rangle,
\end{equation}
where $\bm{u}$ denotes a unit vector with spherical coordinates $(\theta,\varphi)$ and the states are normalized such that $\braket{\bm{u}|\bm{u}}=1$. The expectation values of the spin operators in this state have the simple form
\begin{subequations}\label{eq:Sxyz}
\begin{align}
    \langle \bm{u}|\hat{S}_1|\bm{u}\rangle  &= \tfrac{1}{2}\sin\theta \cos\varphi  \\
    \langle \bm{u}|\hat{S}_2|\bm{u}\rangle  &= \tfrac{1}{2}\sin\theta \sin\varphi  \\
    \langle \bm{u}|\hat{S}_3|\bm{u}\rangle  &= \tfrac{1}{2}\cos\theta.
\end{align}
\end{subequations}
It is then clear that $\{\langle \bm{u}|\hat{S}_i|\bm{u}\rangle\}_{i=1}^3$ are the Cartesian coordinates of a sphere with radius $1/2$. An even more important observation is that $\langle \bm{u}|\hat{S}_i|\bm{u}\rangle$ are orthogonal functions on the sphere:
\begin{equation}
    \int \rd \bm{u}\, \langle \bm{u}|\hat{S}_i|\bm{u}\rangle \langle \bm{u}|\hat{S}_j|\bm{u}\rangle = \frac{1}{6}\delta_{ij},
\end{equation}
where we have defined the integration measure as $\rd \bm{u}=\frac{1}{2\pi}\sin\theta \,\rd\theta\,\rd\varphi$.

We refer to $\langle \bm{u}|\hat{S}_i|\bm{u}\rangle$ as the Q-representation (or Q-function) of the operator $\hat{S}_i$. Likewise the Q-representation of a general operator
\begin{equation}\label{eq:AinIS}
\hat{A}=A_0\hat{\mathcal{I}}+\sum_{i=1}^3A_i\hat{S}_i
\end{equation}
is defined as
\begin{equation}
    A_\text{Q}(\bm{u}) \equiv \langle \bm{u}|\hat{A}|\bm{u}\rangle.
\end{equation}
It is easy to show that this is equivalently written as
\begin{equation*}
    A_\text{Q}(\bm{u}) = \tr[\hat{A}\hat{w}_\text{Q}(\bm{u})], \quad  \hat{w}_\text{Q}(\bm{u}) = \frac{1}{2}\hat{\mathcal{I}} + 2\sum_{i=1}^3 \langle \bm{u}|\hat{S}_i|\bm{u}\rangle  \hat{S_i},
\end{equation*}
where $\hat{w}_\text{Q}$ is the \emph{Stratonovich--Weyl kernel} of the Q-representation.\cite{klimov2009group}
The Q-representation is analogous to the Husimi representation in the nuclear variables.\cite{schleich2011book} It is dual to the P-representation,
\begin{equation*}
    A_\text{P}(\bm{u}) = \tr[\hat{A}\hat{w}_\text{P}(\bm{u})], \quad \hat{w}_\text{P}(\bm{u}) = \frac{1}{2}\hat{\mathcal{I}} + 6\sum_{i=1}^3\langle \bm{u}|\hat{S}_i|\bm{u}\rangle \hat{S_i},
\end{equation*}
which is analogous to the Glauber--Sudarshan representation in the nuclear variables.\cite{schleich2011book} What is meant by `dual' is that any quantum-mechanical trace of a product of operators can be expressed as an integral over a product of Q- and P-symbols as
\begin{equation}\label{eq:trQP}
    \tr[\hat{A}\hat{B}] = \int \rd \bm{u} \, A_\text{Q}(\bm{u})B_\text{P}(\bm{u}) = \int \rd \bm{u} \, A_\text{P}(\bm{u})B_\text{Q}(\bm{u}).
\end{equation}
Most importantly, there is also a W-representation,
\begin{equation*}
    A_\text{W}(\bm{u}) = \tr[\hat{A}\hat{w}_\text{W}(\bm{u})], \quad \hat{w}_\text{W}(\bm{u}) = \frac{1}{2}\hat{\mathcal{I}} + 2\sqrt{3}\sum_{i=1}^3\langle \bm{u}|\hat{S}_i|\bm{u}\rangle \hat{S_i},
\end{equation*}
that is self-dual in the sense that
\begin{equation}\label{eq:trWW}
    \tr[\hat{A}\hat{B}] = \int \rd \bm{u} \, A_\text{W}(\bm{u})B_\text{W}(\bm{u}).
\end{equation}
The W-representation is analogous to the Wigner representation in the nuclear variables. In particular $[\hat{\mathcal{I}}]_\text{W}(\bm{u}) = 1$ and $[\hat{S}_i]_\text{W}(\bm{u})=\sqrt{3}\langle \bm{u}|\hat{S}_i|\bm{u}\rangle$, so that for the general operator in Eq.~\eqref{eq:AinIS} we have
\begin{equation}\label{eq:WA}
    A_\text{W}(\bm{u}) = A_0 + \sqrt{3} \sum_{i=1}^3 A_i \langle \bm{u}|\hat{S}_i|\bm{u}\rangle.
\end{equation}

The introduction of the Stratonovich--Weyl representations Q, P and W for the spin operators
constitutes the major difference between our spin-mapping approach of paper I\cite{runeson2019}
and the spin-mapping models of \Refs{meyer1979spin,Thoss1999mapping,Cotton2015,liu2016}. %
We will employ the Stratonovich--Weyl representation in a similar way for the $N$-level system.

For the $N$-level case, the coherent state will have a more complicated form than in Eq.~\eqref{eq:coherent_state2}, but for the following treatment it is not necessary to write out its explicit expression. We shall denote the generalized coherent state by $|\Omega\rangle$, parametrized by $2N-2$ angles, and leave the details of its construction to Appendix~\ref{sec:app_coh}.
(One can in the following always replace $|\Omega\rangle$ by $|\bm{u}\rangle$ to recover the results of the two-level case.)
For now all we need to know is that $\langle\Omega|\hat{S}_i|\Omega\rangle$ are orthogonal functions such that
\begin{equation}\label{eq:orthogonalityN}
    \int  \rd\Omega\, \langle\Omega|\hat{S}_i|\Omega\rangle \langle \Omega|\hat{S}_j|\Omega\rangle = \frac{1}{2(N+1)}\delta_{ij},
\end{equation}
which is proved in Appendix~\ref{sec:app_aux}. 

It turns out that the Stratonovich--Weyl kernels are remarkably simple to generalize for $N$ levels, as has been shown by Tilma and Nemoto for general $SU(N)$-symmetric coherent states.\cite{Tilma2012} With the choice of normalization in Eq.~\eqref{eq:trSiSj}, the kernels are
\begin{subequations}\label{eq:kernels}
\begin{align}
    \hat{w}_\text{Q}(\Omega) &= \frac{1}{N}\hat{\mathcal{I}} + 2 \sum_{i=1}^{N^2-1}\langle \Omega|\hat{S}_i|\Omega\rangle \hat{S}_i \\
    \hat{w}_\text{P}(\Omega) &= \frac{1}{N}\hat{\mathcal{I}} + 2(N+1) \sum_{i=1}^{N^2-1}\langle \Omega|\hat{S}_i|\Omega\rangle \hat{S}_i \\
    \hat{w}_\text{W}(\Omega) &= \frac{1}{N}\hat{\mathcal{I}} + 2\sqrt{N+1}\sum_{i=1}^{N^2-1}\langle \Omega|\hat{S}_i|\Omega\rangle \hat{S}_i, \label{eq:Wkernel}
\end{align}
\end{subequations}
giving the SW-representations $[\hat{A}]_s(\Omega)=\tr[\hat{A}\hat{w}_s(\Omega)]$ for $s\in\{\text{Q,P,W}\}$.
The readers can easily convince themselves that with this construction, traces of products still obey Eqs.~\eqref{eq:trQP} and~\eqref{eq:trWW} (but with $|\Omega\rangle$ instead of $|\bm{u}\rangle$), as a consequence of Eq.~\eqref{eq:orthogonalityN}.

In the two-level case in paper I,\cite{runeson2019} we interpreted the W-functions of the spin operators as the components of a classical spin vector with magnitude $\sqrt{3}/2$.  Let us define a generalized magnitude as the square-root of
\begin{multline}\label{eq:SWsum}
    \sum_{i=1}^{N^2-1} [\hat{S}_i]_\text{W}(\Omega)^2 = \sum_i [\hat{S}_i]_\text{Q}(\Omega)[\hat{S}_i]_\text{P}(\Omega) \\
    = (N+1)\sum_i \langle \Omega |\hat{S}_i|\Omega\rangle^2 = \frac{N^2-1}{2N},
\end{multline}
where the sum is worked out in Appendix~\ref{sec:app_aux}
and relies on the fact that the $\hat{S}_i$ operators are traceless.
Thus the Casimir invariant in Eq.~\eqref{eq:Casimir} is the generalization of the (squared) spin magnitude, and it is \emph{preserved} by the mapping. 

Many authors have described $\langle \Omega|\hat{S}_i|\Omega\rangle$  as the components of a generalized Bloch vector in $N^2-1$ dimensions.\cite{fano1957,*feynman1957twolevel,*Joos1989,*schlienz1995entanglement,*kimura2003blochvector,hioe1981genGMmat,bertlmann2008qudits} The impact of the W-representation would then be to scale this vector to the length $\tfrac{N^2-1}{2N}$. However, note that not all of its $N^2-1$ components can be independent for $N>2$, since $|\Omega\rangle$ depends only on $2N-2$ spherical variables.
Consequently, only a subset of points on such an $(N^2-1)$-sphere correspond to physical states for $N>2$.\cite{jakobczyk2001blochvector,schirmer2004,kimura2005} We shall therefore not pursue that picture in this paper. %

\begin{figure*}
    \centering
    \includegraphics{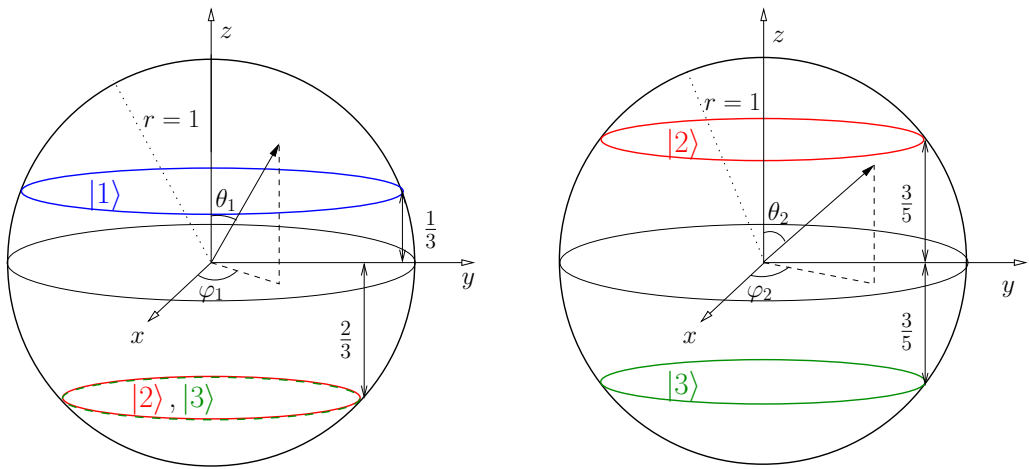}
    \caption{The coherent state $|\Omega\rangle$ of a three-level system is specified by four spherical variables $\theta_1,\varphi_1,\theta_2,\varphi_2$ that we can depict using two spin vectors (here shown in a scale where they have unit magnitude). The coloured circles indicate regions where the W-representation of the system is entirely in one of the diabatic basis states. State $|1\rangle$ corresponds to the blue circle defined by $\cos\theta_1=1/3$ (while $\theta_2$ is arbitrary), state $|2\rangle$ to the red circles defined by $\cos\theta_1=-2/3,\;\cos\theta_2=3/5$, and state $|3\rangle$ to the green circles defined by $\cos\theta_1=-2/3,\;\cos\theta_2=-3/5$. This picture generalizes the two-level case in Fig.~2 of paper I.\cite{runeson2019} For a general $N$-level system one would have $N-1$ spheres.}
    \label{fig:spheres}
\end{figure*}

A more natural picture would be to think of the $2N-2$ spherical variables $\{\theta_n,\varphi_n\}_{n=1}^{N-1}$ of $|\Omega\rangle$ as the orientations of $N-1$ spin-$\thalf$ vectors (their interpretation as spins is explained in Appendix~\ref{sec:app_spin1}). Fig.~\ref{fig:spheres} shows an example for $N=3$. In the two-level case, we saw in paper I\cite{runeson2019} that the orientations corresponding to single basis states were found at the poles in the Q-representation, but at ``polar circles'' with fixed $\theta$ in the W-representation.  In Appendix~\ref{sec:app_coh} it is worked out that the W-representation of the basis states can be represented by polar circles also for $N>2$, but on different latitudes from the $N=2$ case. This is in contrast to the Q-representation, where these circles would be replaced by points at the respective poles.

While this picture is instructive as a generalization to Fig.~2 in paper I,\cite{runeson2019} it does not help us find the equations of motion of the system.
To describe the dynamics in a simple way, we shall therefore from now on switch to a Cartesian representation. This will also reveal the link between the generalized spin mapping and the MMST mapping.

\subsection{Dynamics in Cartesian variables}\label{sec:dynamics}
An alternative to using spherical variables is to write the coherent states in terms of complex coefficients $\{c_n\}$:
\begin{equation}
    |\Omega\rangle = c_1 |1\rangle + c_2 |2\rangle + \dots + c_N|N\rangle.
\end{equation}
Given the constraint $\sum_n |c_n|^2=1$ and an arbitrary choice of global phase, the $\{c_n\}$ have $2N-2$ real degrees of freedom. 
Starting as usual with the two-level coherent state, $\ket{\bm{u}}=c_1\ket{1}+c_2\ket{2}$, the orthogonal functions in Eq.~\eqref{eq:Sxyz} take the form
\begin{subequations}\label{eq:Sc1c2}
\begin{align}
    \langle \bm{u}|\hat{S}_1|\bm{u}\rangle &= \thalf(c_1^*c_2 + c_2^*c_1) = \Re(c_1^*c_2) \\
    \langle \bm{u}|\hat{S}_2|\bm{u}\rangle &= -\tfrac{\ii}{2}(c_1^*c_2-c_2^*c_1)=\Im(c_1^*c_2) \\
    \langle \bm{u}|\hat{S}_3|\bm{u}\rangle &= \tfrac{1}{2} (|c_1|^2-|c_2|^2).
\end{align}
\end{subequations}
Let us insert these into the W-representation of an arbitrary operator in Eq.~\eqref{eq:WA}:
\begin{multline}
    A_\text{W} = A_0 + \sqrt{3}\left[A_1 \Re(c_1^*c_2) + A_2 \Im(c_1^*c_2) \right. \\
    \left. + A_3 \thalf (|c_1|^2-|c_2|^2)   \right].
\end{multline}
We now introduce the Cartesian variables $X_n$ and $P_n$ via $3^{1/4}c_n \equiv \frac{1}{\sqrt{2}}(X_n+\ii P_n)$. Because of $\sum_n|c_n|^2=1$, these are constrained to a sphere with squared radius
\begin{equation}\label{eq:R2}
    R^2=X_1^2+P_1^2+X_2^2+P_2^2=2\sqrt{3}.
\end{equation}
The general W-representation becomes
\begin{multline}
    A_\text{W}(X,P) = A_0 + \thalf\left[A_1(X_1 X_2+P_1P_2)  \right. \\ 
    \left. + A_2(X_1 P_2-X_2 P_1) + A_3\thalf(X_1^2+P_1^2-X_2^2-P_2^2) \right],
\end{multline}
for which a particularly important case is the two-level Hamiltonian (which we choose to be real),
\begin{multline}
    H_\text{W}(X,P)= \frac{p^2}{2m} + \frac{V_1(x)+V_2(x)}{2} + \Delta(x)(X_1X_2+P_1P_2) \\ 
    +\frac{V_1(x)-V_2(x)}{2} \thalf (X_1^2+P_1^2-X_2^2-P_2^2).
\end{multline}
With the use of Eq.~\eqref{eq:R2}, this can also be written
\begin{multline}
    H_\text{W}(X,P)= \frac{p^2}{2m}  +\sum_{n=1}^2 V_n(x)\thalf(X_n^2+P_n^2-\gamma) \\
        + \Delta(x)(X_1X_2+P_1P_2),
\end{multline}
where $\gamma=\sqrt{3}-1$ is called the \emph{zero-point energy parameter}.
(Note that some authors use an alternative convention used for $\gamma$, which is half the value of ours.)
This is similar to the Hamiltonian derived by Meyer, Miller, Stock and Thoss,\cite{Meyer1979nonadiabatic,Stock1997mapping} the difference being that $\gamma=1$ in their formulation. This value emerged from a Langer correction in Meyer and Miller's formulation, and from the commutation relations of harmonic-oscillator operators in the formulation by Stock and Thoss.
Stock and M\"{u}ller have however observed that decreasing the value of $\gamma$ often gives more accurate results,\cite{muller1999a,Mueller1999pyrazine} and suggested choosing $\gamma\approx \thalf$ as a general rule of thumb.\cite{Stock1995meanfieldII,Stock2005nonadiabatic}
The particular value $\gamma=\sqrt{3}-1\approx 0.732$ has previously been proposed by Cotton and Miller\cite{Cotton2013b} and appears naturally from the Stratonovich--Weyl formalism for a two-level system mapped to a spin-$\thalf$.\cite{runeson2019} We will now derive the value of $\gamma$ from the Stratonovich--Weyl representation for a general $N$-level system.

Again we focus our attention on the W-representation and simply state the Q- and P-versions at the end of the section. By construction, it is clear that
\begin{equation}
    \langle \Omega|\hat{S}_i|\Omega\rangle = \sum_{n,m=1}^N \langle n|\hat{S}_i|m\rangle c_n^*c_m.
\end{equation}
Given the kernel in Eq.~\eqref{eq:Wkernel}, the W-representation of $\hat{S}_i$ is 
\begin{equation}
    [\hat{S}_i]_\text{W}(\Omega) = \sqrt{N+1}\langle \Omega|\hat{S}_i|\Omega\rangle =  \sum_{n,m=1}^N\langle n|\hat{S}_i|m\rangle \sqrt{N+1} c_n^*c_m.
\end{equation}
Let us therefore introduce $X_n$ and $P_n$ through
\begin{equation}\label{eq:XPdef}
    (N+1)^{1/4}c_n = \frac{X_n+\ii P_n}{\sqrt{2}},
\end{equation}
from which follows that $X_n$ and $P_n$ are constrained to a hypersphere with squared radius
\begin{equation}\label{eq:R2N}
    R^2 \equiv \sum_{n=1}^N (X_n^2 + P_n^2) = 2\sqrt{N+1}.
\end{equation}
Note that when $N=2$, this result reduces to \eqn{R2}.
To find the value of $\gamma$ in the Hamiltonian, consider its decomposition in basis matrices according to Eq.~\eqref{eq:Sbasis}
with $H_0=\frac{p^2}{2m} + \bar{V}(x)$, where $\bar{V}(x)= \frac{1}{N}\sum_{n=1}^N V_n(x)$, and $H_i=2\,\tr[\hat{V}(x)\hat{S}_i]$.
It is clear from Eq.~\eqref{eq:Wkernel} that the W-representation scales all matrices except for the identity by $\sqrt{N+1}$. This means that
\begin{multline}
    H_\text{W} = \frac{p^2}{2m}+ \bar{V}(x) + \\ \sqrt{N+1}\left(\sum_{n=1}^N(V_n(x)-\bar{V}(x))|c_n|^2
    +\sum_{n\neq m}V_{nm}(x) c_n^*c_m \right).
\end{multline}
By inserting \eqn{XPdef}, we finally recover the MMST-form of the Hamiltonian (which is again chosen to be real):
\begin{multline}\label{eq:HW}
    H_\text{W}(X,P) = \frac{p^2}{2m} + \sum_n V_n(x) \tfrac{1}{2}(X_n^2+P_n^2-\gamma) \\
        + \sum_{n>m}V_{nm}(x)(X_n X_m + P_n P_m),
\end{multline}
with
\begin{equation}\label{eq:gamma}
    \gamma=\frac{2}{N}(\sqrt{N+1}-1).
\end{equation}
This is always smaller than the value $\gamma=1$ that Stock and Thoss  obtained from the commutation relations $[\hat{X}_n,\hat{P}_n]=1$ of the harmonic oscillator operators $\hat{X}_n,\hat{P}_n$ of the extended mapping space. Note that we have not invoked such an extended space, but instead derived Eq.~\eqref{eq:gamma} purely based on the commutation relations of the original problem that enter through the Casimir invariant.
The new value of $\gamma$ as well as the constraint in Eq.~\eqref{eq:R2N} are both minor modifications to any code that uses MMST mapping or Ehrenfest dynamics, but lead to significant improvements in accuracy, as we shall show in Sec.~\ref{sec:FMO}. 

The derivation holds under the assumption that the symmetry group of the subsystem is $SU(N)$. This means for example that we assume that no level is decoupled from all the others, since that would lead to a reduction of the symmetry group and change the value of $\gamma$. A treatment in terms of time-dependent symmetry groups would be much more involved\cite{zhang1995quantum} and is outside the scope of this paper. In the case of $N=1$ there is only one basis matrix (the identity of shape $1\times1$) so that $H_\text{W}=H_0$ and $\gamma$ does not appear, which recovers the standard single-surface Born--Oppenheimer Hamiltonian.

The same analysis can be done for the Q- and the P-functions, and the results for $R^2$ and $\gamma$ are summarized in Table~\ref{tab:R2gamma}.
Note that $\gamma$ is independent of $N$ for the Q- and the P-representations, but decreases with $N$ in the W-representation, and that all quantities coincide with our previous results\cite{runeson2019} for $N=2$.

What we have presented is a major step forward from the previously suggested spin-mapping approaches,\cite{meyer1979spin,Cotton2015,liu2016}
which did not recover the MMST Hamiltonian and did not reduce to the exact quantum dynamics for an isolated subsystem. One could even say that the generalized spin mapping is a more natural derivation of the MMST Hamiltonian, as it requires no extended phase space and directly gives a $\gamma$ closer to those found optimal in numerical simulations. In the rest of this section we will give two further reasons for this point of view. %

\begin{table}%
    \centering
    \caption{Formulas for the squared $XP$-radius $R^2$ and the zero-point energy parameter $\gamma$ for general $N$.}
    \label{tab:R2gamma}
    \begin{ruledtabular}
    \begin{tabular}{ccc}
        $s$  &  $R^2$ & $\gamma$ \\  \hline
        Q & 2 & 0 \\ 
        W & $2\sqrt{N+1}$ & $\frac{2}{N}(\sqrt{N+1}-1)$ \\
        P & $2(N+1)$ & 2 \\
    \end{tabular}
    \end{ruledtabular}
\end{table}

The equations of motion that correspond to the Hamiltonian in Eq.~\eqref{eq:HW} look identical to those of the MMST-mapping:
\begin{subequations}\label{eq:eom_all}
\begin{align}
    \dot{X}_n &= \sum_{m=1}^N V_{nm} P_m \\ 
    \dot{P}_n &= - \sum_{m=1}^N V_{nm} X_m \\
    \dot{x} &= p/m \\
        \dot{p} &= - \sum_{n=1}^N \pd{V_n}{x} \frac{1}{2}(X_n^2+P_n^2-\gamma) \nonumber\\
        &\qquad - \sum_{n>m}\pd{V_{nm}}{x}(X_n X_m + P_n P_m),
\end{align}
\end{subequations}
but in contrast to the MMST mapping, our $X_n$ and $P_n$ are constrained to a hypersphere with squared radius $R^2$ (see Table~\ref{tab:R2gamma}) and so the ensemble dynamics will be subtly different. Each trajectory exactly preserves $R^2$, %
so that there is \emph{no leakage} out of the mapping space, thereby solving a problem of the original MMST mapping.
Like in the MMST dynamics, the classical equations of motion for the $X_n$ and $P_n$ variables exactly correspond to the electronic Schr\"{o}dinger equation for the uncoupled system.

In literature it is common to separate $\hat{V}(x)$ into a state-independent and a state-dependent part, and the particular choice of splitting can influence the results of some methods (such as LSC-IVR and SQC).
Typically authors suggest to separate the traced and traceless parts \cite{Kelly2012mapping,saller2019jcp,Hsieh2012FBTS}
although other choices have also been used.\cite{Wang1999mapping,mapping}
In our approach, the weights of the potential-energy surfaces always sum up to one:
\begin{equation}
    \sum_{n=1}^N \thalf(X_n^2+P_n^2-\gamma) = 1,
\end{equation}
which means that it is \emph{independent} of such a splitting.

\subsection{Correlation functions}\label{sec:correlation}
We will now use the results of Secs.~\ref{sec:SW}--\ref{sec:dynamics} to approximate the correlation function
\begin{equation}
    C_{AB}(t) = \mathrm{Tr}[\hat{\rho}\hat{A}(0)\hat{B}(t)],
\end{equation}
where capitalized $\Tr$ means a trace over both electronic and nuclear states.
The trace over the electronic degrees of freedom can be written as integrals of Stratonovich--Weyl functions (see Eqs.~\eqref{eq:trQP} and~\eqref{eq:trWW}).
Likewise, the trace over the nuclei can be expressed in terms of the Wigner distribution
\begin{equation}\label{eq:rhonucW}
    \rho_\text{nuc}(x,p) = \int \eu{\ii py}\left\langle x-\frac{y}{2}\right|\hat{\rho}_\text{nuc} \left|x+\frac{y}{2}\right\rangle \rd y,
\end{equation}
and we choose initial conditions such that $\hat{\rho}=\hat{\rho}_\text{nuc}\otimes \hat{\mathcal{I}}$. (The initial electronic state is defined by $\hat{A}$.) %

Then the correlation function can be exactly written as
\begin{equation}
    C_{AB}(t) = N\langle A_{s}(\mat{X},\mat{P}) [\hat{B}(t)]_{\bar{s}}(\mat{X},\mat{P}) \rangle,
\end{equation}
where $s\in\{\text{Q,\;P,\;W}\}$, $\bar{s}$ is the dual of $s$, and
\begin{equation*}
    \braket{\cdots} = \frac{\int \rd x\,\rd p \, \rd \mat{X}\, \rd\mat{P} \cdots \delta(\mat{X}^2+\mat{P}^2-R_s^2) \rho_\text{nuc}(x,p)}{\int \rd x\,\rd p \, \rd \mat{X}\, \rd\mat{P}\,\delta(\mat{X}^2+\mat{P}^2-R_s^2)\rho_\text{nuc}(x,p)}.
\end{equation*}
The additional factor of $N$ appears because we have defined $\Tr[\hat{\rho}]=N$ but $\langle [\hat{\rho}]_s \rangle = 1$, the subscript $s$ on $R_s$ specifies which radius to use from Table~\ref{tab:R2gamma}, and we used the shorthand notation $X^2\equiv \sum_{n=1}^N X_n^2$ and similarly for $P^2$. 

We now propose to approximate the correlation function by
\begin{equation} \label{eq:CAB}
    C_{AB}(t) \approx N \langle A_{s}(\mat{X},\mat{P}) B_{\bar{s}}(\mat{X}(t),\mat{P}(t))\rangle,
\end{equation}
where the dynamics is driven by the Hamiltonian $H_s$ in the $s$-representation. This formula is the multi-level generalization of the quasiclassical spin-mapping method in paper I.\cite{runeson2019} Note that for $s=\text{W}$, we are using classical Wigner dynamics in both the nuclear and the electronic degrees of freedom.
In paper I,\cite{runeson2019} we showed how the dynamics follows from approximating the time derivative of $\hat{B}$ with a Poisson bracket, similar to the approximation used in PBME.\cite{Kelly2012mapping}

Typically we will be interested in population transfer from a state $n$ to a state $m$, i.e.\ $\hat{A}=|n\rangle\langle n|$ and $\hat{B}=|m\rangle\langle m|$. The corresponding Stratonovich--Weyl functions are the population observables
\begin{subequations}
\begin{align}
    [|n\rangle\langle n|]_s &= \thalf(X_n^2+ P_n^2-\gamma_s) \label{eq:Aop} \\
    [|m\rangle\langle m|]_{\bar{s}} &= \thalf \left(\frac{R^2_{\bar{s}}}{R^2_s}(X_m^2+P_m^2)-\gamma_{\bar{s}}\right),\label{eq:Bop}
\end{align}
\end{subequations}
where we use subscripts on $\gamma$ and $R^2$ to distinguish between the $s$- and the $\bar{s}$-symbols. The factor $R_{\bar{s}}^2/R_s^2$ appears when $s=\text{Q}$ or P, because the $X_n$ and $P_n$ variables are sampled from a hypersphere with radius $R_s$ but measured on a sphere with radius $R_{\bar{s}}$.
In the symmetric case of $s=\bar{s}=\text{W}$, Eqs.~\eqref{eq:Aop} and~\eqref{eq:Bop} reduce to the same expression. 
It is however no more difficult to calculate off-diagonal elements of the density matrix, for example:
\begin{subequations}
\begin{align}
    [|n\rangle\langle m| + |m\rangle\langle n|]_{s} = X_nX_m+P_nP_m \\
    \ii[|n\rangle\langle m| - |m\rangle\langle n|]_{s} = X_nP_m-P_nX_m,
\end{align}
\end{subequations}
and the $\bar{s}$ symbols are again obtained by multiplying with $R_{\bar{s}}^2/R_s^2$.

We thus have the alternative of calculating the correlation function in a symmetric way, meaning ($s,\bar{s})$=(W,W), or in an asymmetric way, that is ($s,\bar{s})$=(Q,P) or (P,Q). In paper I\cite{runeson2019} we saw that (Q,P) and (W,W) both gave accurate results for a wide range of spin-boson models, while (P,Q) was always less accurate. After running tests on further systems we have observed that 
(Q,P) is not always so reliable but that
the symmetric definition (W,W) is the most robust. This is confirmed by the results we show in Sec.~\ref{sec:FMO} and in the Supplementary Material.%

The initial distribution of the mapping variables that follows from the Stratonovich--Weyl formalism is a uniform distribution over the sphere $\mat{X}^2+\mat{P}^2=R_s^2$. We will call this  \emph{full-sphere} initial conditions. Note that it gives the results of all $n\to m$ transitions in a single simulation. The sampling of the distribution $\delta(X^2+P^2-R_s^2)$ is easy to implement in practice by drawing $\{X_n,P_n\}_{n=1}^N$ from a standard normal distribution and rescaling them with a common factor so that $X^2+P^2=R_s^2$.

Previous authors in the mapping community have also used an approximation called \emph{focused} initial conditions\cite{Bonella2003mapping,Bonella2005LANDmap,muller1999a,Mueller1999pyrazine,Kim2008Liouville,huo2012focus,Hsieh2013FBTS} in which the initial distribution only includes points that directly correspond to coherent states that diagonalize $\hat{A}$, rather than by weighting as in Eq.~\eqref{eq:CAB}. In the case of $\hat{A}=|n\rangle\langle n|$, this means that $\thalf(X_n^2+P_n^2-\gamma)=1$ while $\thalf(X_k^2+P_k^2-\gamma)=0$ for all $k\neq n$, or equivalently
\begin{equation}
    X_k=r_k \cos\phi_k, \quad P_k = r_k \sin\phi_k,
\end{equation}
with $r_{k=n}=\sqrt{2+\gamma},~r_{k\neq n}=\sqrt{\gamma}$ and uniformly sampled $\phi_k\in[0,2\pi)$. This can be interpreted as sampling from the ``polar circles'' in Fig.~\ref{fig:spheres}. In the correlation function, $A(X,P)=1$ by construction, so that
\begin{equation}
    C_{AB}(t)\approx \langle B_s(X(t),P(t))\rangle_\text{foc}
\end{equation}
where
\begin{equation*}
    \braket{...}_\text{foc}= \frac{\int \rd x\,\rd p \, \rd \mat{X}\, \rd\mat{P} \cdots \rho_\text{foc}(X,P) \rho_\text{nuc}(x,p)}{\int \rd x\,\rd p \, \rd \mat{X}\, \rd\mat{P}\, \rho_\text{foc}(X,P)\rho_\text{nuc}(x,p)},
\end{equation*}
uses the focused distribution
\begin{equation}
    \rho_\text{foc}(X,P)= \delta(X_n^2+P_n^2-\gamma-1)\prod_{k\neq n} \delta(X_k^2+P_k^2-\gamma),
\end{equation}
and trajectories are defined according to $H_s$.
Note that for focused methods the observable $B_s(X(t),P(t))$ must be calculated with the \emph{same} index $s$ as the Hamiltonian, in contrast with methods that use full-sphere initial conditions.\footnote{This is necessary to get the correct value at time zero.} It is also possible to define focused initial conditions when starting from off-diagonal elements of the density matrix, as explained in paper I.\cite{runeson2019}

This prescription is the same as that of M\"{u}ller and Stock,\cite{Mueller1999pyrazine} apart from the value of $\gamma$ in Eq.~\eqref{eq:gamma}, which we derived instead of treating it as a free parameter. Our $\gamma$ decreases with $N$, but the convergence to zero is so slow that only with $N\geq 360$ does it reach $\gamma\leq 0.1$. It is interesting to note that the limit $N\to 0$ gives the standard MMST-value $\gamma=1$, and that the spin-mapping value is therefore somewhere between that of MMST and the Ehrenfest value $\gamma=0$. A comparison of numerical values for $\gamma$ is given in Table~\ref{tab:gamma}. 
One might ask if the $N\to\infty$ limit of focused W would be equivalent to the Ehrenfest method, which is the same as focused Q. Bearing in mind that $R_\text{W}$ diverges as $N\to\infty$, while $\gamma\to 0$, this is still an open question.

Finally note that if one applies focused initial conditions to a problem without inter-state couplings, the prescription reduces to Born--Oppenheimer dynamics on the initial state.

\begin{table}%
    \centering
    \caption{Examples of numerical values of the zero-point energy parameter $\gamma$ in the W-representation for different number of levels $N$, evaluated with Eq.~\eqref{eq:gamma}. For comparison, the lowest rows show the corresponding values in other methods. (Note that there are two definitions of $\gamma$ in literature, of which one is a factor of 2 smaller than in our definition.)}\label{tab:gamma}
    \begin{ruledtabular}
    \begin{tabular}{cr}
        $N$ & $\gamma$  \\ \hline
        2   & $\sqrt{3}-1\approx 0.732$ \\
        3   & $2/3 \approx 0.667$ \\
        7   & $0.522$ \\
        8  & $1/2 = 0.500$\\
        100 & $0.181$ \\        \hline 
        Standard MMST & $1$ \\
        Ehrenfest & $0$ \\ 
        SQC with square windows\cite{Cotton2013b} & $\sqrt{3}-1\approx 0.732$  \\
        SQC with triangular windows\cite{Cotton2016,Cotton2019manystates} & $2/3\approx 0.667$  \\  
    \end{tabular}
    \end{ruledtabular}
    \label{tab:my_label}
\end{table}

\section{Application to the Fenna--Matthews--Olson model}\label{sec:FMO}
We have tested the theory of Sec.~\ref{sec:theory} on a seven-level model of the Fenna--Matthews--Olson (FMO) complex, which is a well-studied light-harvesting pigment-protein complex found in green sulphur bacteria.\cite{fenna1975chlorophyll} Each diabatic state represents an exciton localized on one of the sites. This is a challenging benchmark problem for electronically nonadiabatic dynamics and allows our method to be compared with other mapping approaches,\cite{Huo2011densitymatrix,cotton2016lightharvest,saller2020faraday,Cotton2019manystates} as well as to numerically exact results obtained via the hierarchical equation of motion (HEOM) approach.\cite{ishizaki2009unified,ishizaki2009FMO,ishizaki2010PCCP,zhu2011FMO,dattani2015fmo} 

\subsection{Model description}\label{sec:FMOmodel}
The model Hamiltonian is of the subsystem-bath type:
\begin{equation}
    \hat{H} = \hat{H}_\text{s} + \hat{H}_\text{b} + \hat{H}_\text{sb},
\end{equation}
where the subsystem Hamiltonian in a diabatic basis is given in units of cm$^{-1}$ as\cite{adolphs2006fmo}
\begin{equation}
    \hat{H}_\text{s} = \begin{pmatrix} 
    12410 & -87.7 & 5.5 & -5.9 & 6.7 & -13.7 & -9.9 \\ 
    -87.7 & 12530 & 30.8 & 8.2 & 0.7 & 11.8 & 4.3 \\
    5.5 & 30.8 & 12210 & -53.5 & -2.2 & -9.6 & 6.0 \\
    -5.9 & 8.2 & -53.5 & 12320 & -70.7 & -17.0 & -63.3 \\
    6.7 & 0.7 & -2.2 & -70.7 & 12480 & 81.1 & -1.3 \\
    -13.7 & 11.8 & -9.6 & -17.0 & 81.1 & 12630 & 39.7 \\
    -9.9 & 4.3 & 6.0 & -63.3 & -1.3 & 39.7 & 12440
    \end{pmatrix}.
\end{equation}
Each level is coupled to its own bath of $F$ nuclear modes with unit mass and frequencies $\omega_j$, so that the total bath Hamiltonian is
\begin{equation}
    \hat{H}_\text{b} = \sum_{n=1}^N \sum_{j=1}^F  \left( \tfrac{1}{2} p_{j,n}^2  + \tfrac{1}{2}\omega_j^2 x_{j,n}^2\right)\hat{\mathcal{I}}.%
\end{equation}
The baths of the different sites have identical frequencies and are not directly coupled to each other. The system-bath coupling is in turn
\begin{equation}
    \hat{H}_\text{sb} = \sum_{n=1}^N \sum_{j=1}^F \xi_{j}x_{j,n}|n\rangle\langle n|,
\end{equation}
with coupling coefficients $\xi_{j}$. 
The frequencies are distributed according to a Debye spectral density
\begin{equation}
    J(\omega) = 2\lambda \frac{\omega \omega_\text{c}}{\omega^2 + \omega_\text{c}^2},
\end{equation}
where $\omega_\text{c}$ is the characteristic frequency of the bath ($\tau_\text{c}=\omega_\text{c}^{-1}$ is its corresponding timescale) and $\lambda$ is the reorganization energy. In accordance with previous work,\cite{ishizaki2009FMO,ishizaki2010PCCP,zhu2011FMO,Cotton2019manystates,saller2020faraday} we used $\lambda=35$\;cm$^{-1}$ in all our simulations. We used a discretization of the bath with $F=60$ modes per site (in total 420 modes), according to the discretization scheme in Ref.~\onlinecite{Craig2007condensed}. The discretization also determines the coupling coefficients $\xi_j$.

To initialize the nuclear bath, we sampled the Wigner distribution in Eq.~\eqref{eq:rhonucW}, which is explicitly
\begin{equation}
    \rho_\text{nuc}(x,p)=\prod_{n=1}^N \prod_{j=1}^F \frac{\alpha_j}{\pi} 
    \exp\left[-\alpha_j\left(\frac{p_{j,n}^2}{\omega_j}+\omega_jx_{j,n}^2\right)\right],
\end{equation}
where $\alpha_j = \tanh\tfrac{\beta\omega_j}{2}$.

Each simulation was run with timestep 1\,fs and $10^6$ trajectories to guarantee convergence, although we point out that it was possible to observe the trend of each line already with $10^3$ trajectories.

\subsection{Population dynamics}\label{sec:FMOresults}
We have tested our theory on the FMO model using the same parameters as in the original paper by Ishizaki and Fleming.\cite{ishizaki2009FMO}
In each figure we will compare six methods, where upper panels correspond to full-sphere initial conditions and lower panels to focused initial conditions (these are defined in Sec.~\ref{sec:correlation}). Within each row $\gamma$ increases from left to right, so that left panels display methods with $\gamma=0$ (the lowest possible), middle panels the W-value of $\gamma$ from Eq.~\eqref{eq:gamma}, and right panels the large-$\gamma$ case. In the upper right panel this is the P-value, while in the lower right panel we show the standard MMST-value $\gamma=1$ for the focused method (the P-value would typically be worse).
Note that the focused method with $\gamma=0$ (lower left panels) is identical to Ehrenfest dynamics.  

\begin{figure*}%
\includegraphics{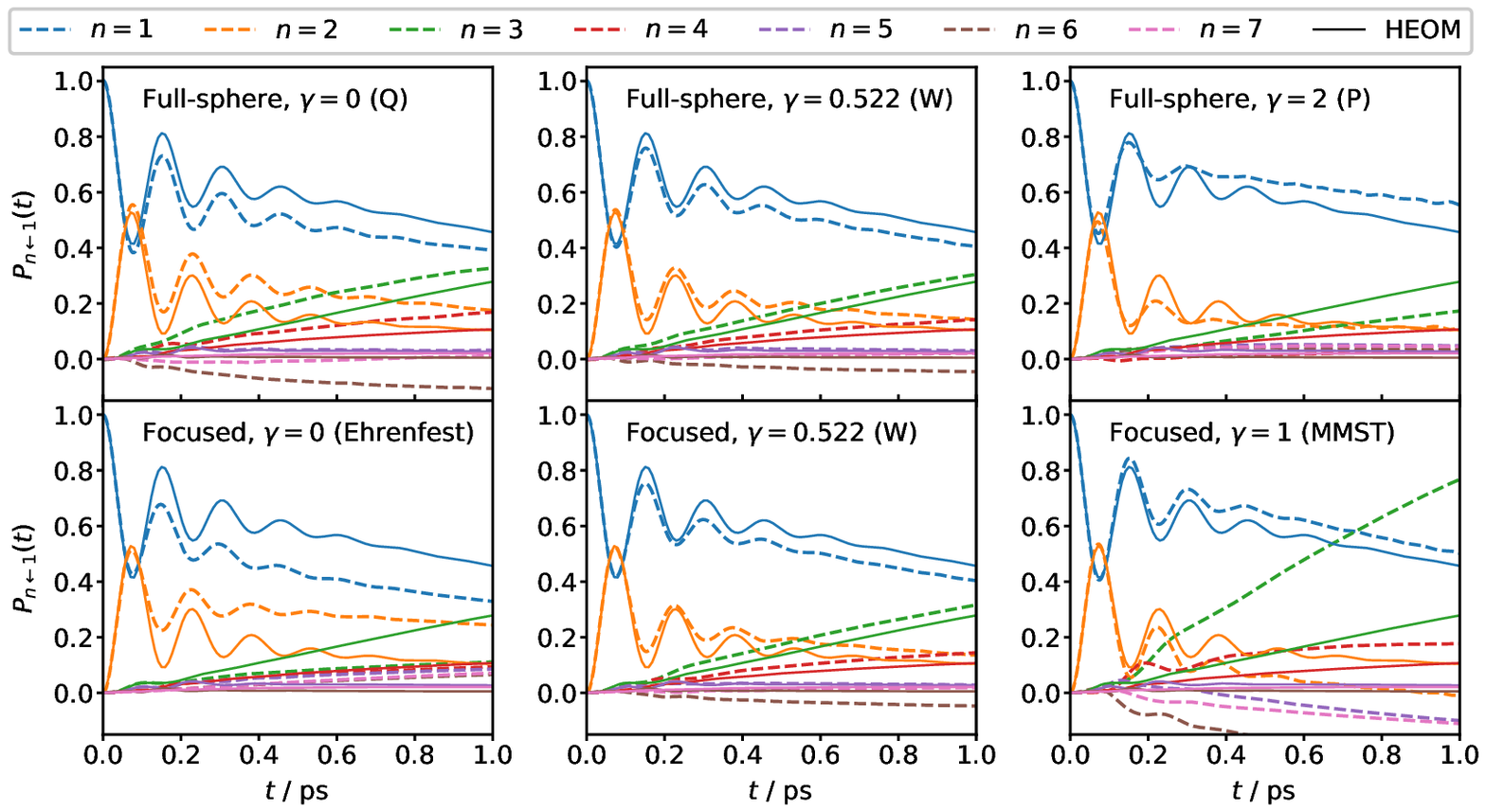}
\caption{Results (dashed lines) for a 7-state FMO model at $T=\SI{77}{K}$ with $\tau_\text{c}=\SI{50}{fs}$ ($\omega_\text{c}=\SI{106.14}{cm^{-1}}$), starting from state~1. Solid lines show numerically exact HEOM results.\cite{ishizaki2009FMO} The middle column uses the W-value of $\gamma$ from Eq.~\eqref{eq:gamma} that is derived in this paper. We encourage the reader to compare these results with SQC (see Fig.~10a of Ref.~\onlinecite{Cotton2019manystates} and Fig.~3 of Ref.~\onlinecite{cotton2016lightharvest}), traceless MMST (Figs.~2--3 of Ref.~\onlinecite{saller2020faraday}), and PLDM (Fig.~3 of Ref.~\onlinecite{Huo2011densitymatrix} and Fig.~6 of Ref.~\onlinecite{Mandal2018quasidiabatic}).}\label{fig:A1_short}
\end{figure*}

Fig.~\ref{fig:A1_short} shows the results for low temperature ($T=\SI{77}{K}$) and a fast bath ($\tau_\text{c}=\SI{50}{fs}$), which is the hardest of the model problems since it has the strongest quantum effects. It is clear from each row that dynamics using $\gamma$ derived in the W-representation is generally more accurate than the other cases. For W, the difference between using full-sphere or focused initial conditions is negligible. Typically the focused methods converge with an order of magnitude fewer trajectories than the full-sphere methods.

\begin{figure*}%
\includegraphics{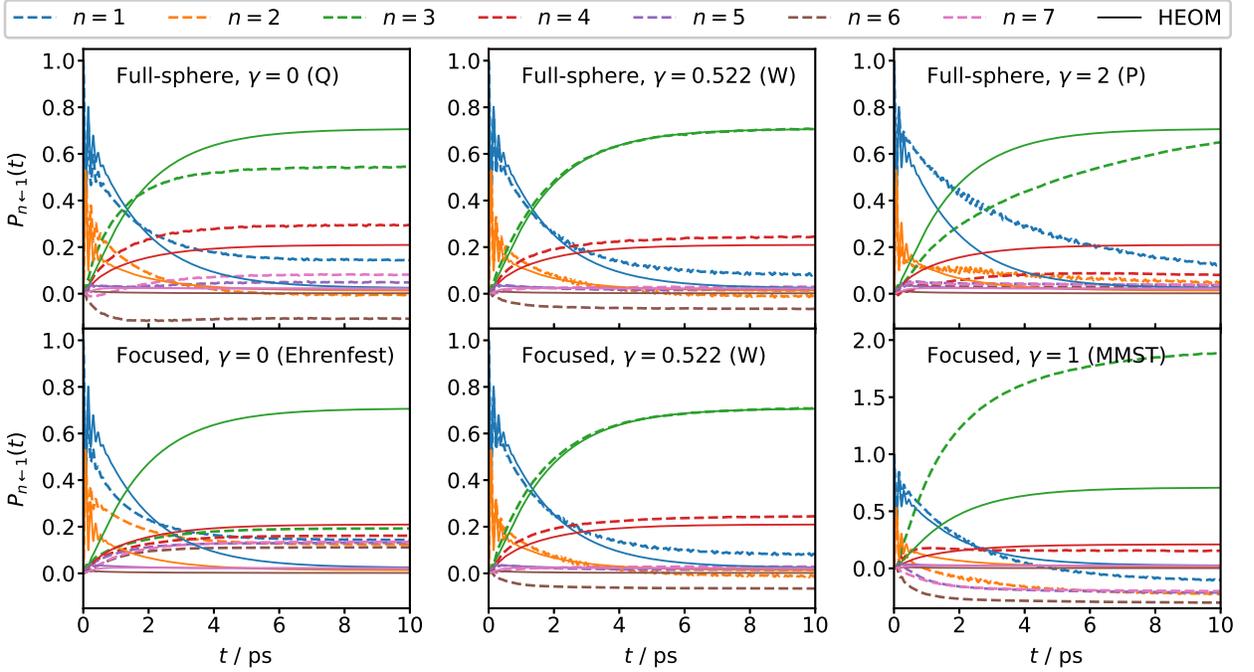}
    \caption{Long-time results for the FMO model in Fig.~\ref{fig:A1_short} ($T=\SI{77}{K}$, $\tau_\text{c}=\SI{50}{fs}$, initial state 1). Solid lines show numerically exact HEOM results.\cite{dattani2015fmo} The reader may compare this figure with traceless MMST (Fig.~6 of Ref.~\onlinecite{saller2020faraday}).}\label{fig:A1_long}
\end{figure*}

These observations become even clearer when looking at the long-time limit of the same model in Fig.~\ref{fig:A1_long}. Again the W-value of $\gamma$ derived in Eq.~\eqref{eq:gamma} is the most accurate for the final populations, while Ehrenfest and focused MMST are very unreliable. The W-methods may still predict unphysical negative populations, but the absolute error is still typically smaller than in the other methods (and note that such negative populations are also possible in other mapping approaches\cite{saller2020faraday,sun1998semiclassical,Kim2008Liouville,Huo2011densitymatrix,Hsieh2012FBTS}).

\begin{figure*}%
\includegraphics{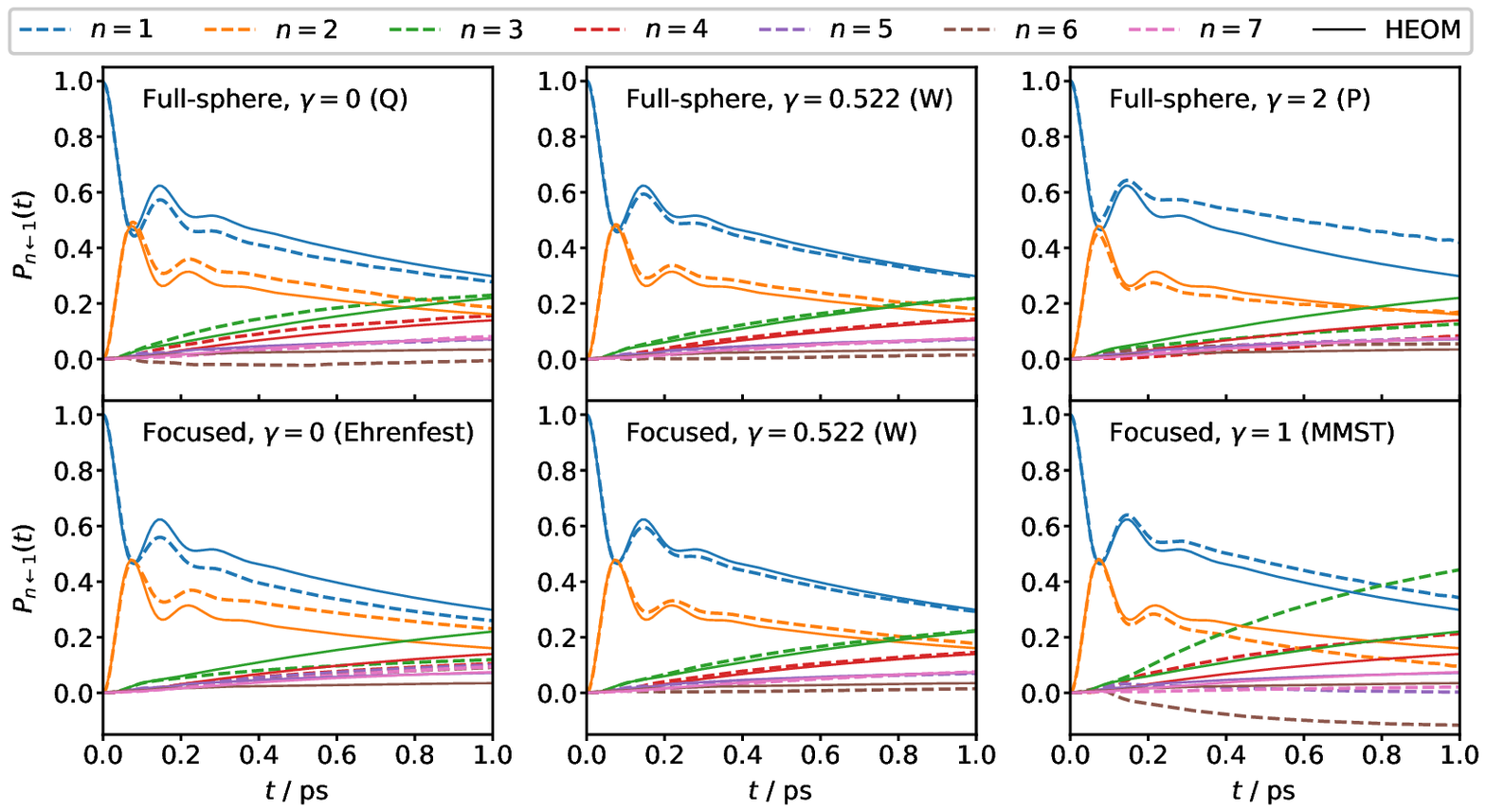}
\caption{Results for the FMO model at $T=300$\,K with $\tau_\text{c}=50$ fs, starting from state 1. Solid lines show numerically exact HEOM results.\cite{ishizaki2009FMO} The reader may compare this figure with SQC (Fig.~10c of Ref.~\onlinecite{Cotton2019manystates}) and traceless MMST (Fig.~4 of Ref.~\onlinecite{saller2020faraday}).}\label{fig:B1_short}
\end{figure*}

Finally we show the results for a higher temperature ($T=\SI{300}{K}$) in Fig.~\ref{fig:B1_short}. This problem is not as hard as the previous, in the sense that all methods have decent accuracy, but it is still clear that W is the most accurate. 
In the Supplementary Material we show the case of starting from state 6 instead of 1, as well as the case of a slow bath ($\tau_\text{c}=\SI{166}{fs}$), and they give further weight to our conclusions.

It should be noted that the middle column (W) results shown here  clearly outperform both PBME,\cite{kelly2011fmo} LSC-IVR\cite{tao2010FMO} and Ehrenfest dynamics.
The W results are of similar accuracy to other state-of-the-art mapping approaches such as SQC\cite{Cotton2019manystates} and traceless MMST.\cite{saller2020faraday} They are slightly less accurate than PLDM for this system,\cite{Huo2011densitymatrix,Mandal2018quasidiabatic} but the improvement of PLDM compared to linearized MMST suggests that a (future) partially linearized version of the spin-mapping method might perform even better.
As discussed in Sec.~\ref{sec:dynamics}, the W approach has the advantages compared to MMST-based methods that it does not require choosing a window function, it has no leakage from the mapping space, and it is independent of the splitting of the potential-energy matrix.%

\subsection{Bipartite entanglement}
We now turn to the problem of calculating off-diagonal elements of the density matrix. A quantity for which numerically exact benchmarks exist is the bipartite entanglement (sometimes called concurrence) between state $n$ and $m$, defined as $2|\rho_{nm}(t)|$, after an initial excitation to one of the states. Here $\rho_{nm}(t)$ denotes elements of the reduced density matrix of the subsystem.
More specifically, we compute the correlation functions $C_{AS^+}$ and $C_{AS^-}$ with $\hat{A}=|k\rangle\langle k|$ being the initial state and $\hat{S}^+ =|n\rangle \langle m|+|m\rangle\langle n|$ for $2\,\Re[\rho_{nm}]$ and $\hat{S}^-=i(|n\rangle\langle m|-|m\rangle\langle n|)$ for $2\,\Im[\rho_{nm}]$. The time-dependent concurrence is then given by
\begin{equation}
    2|\rho_{nm}(t)| = \sqrt{C_{AS^+}^2(t)+C_{AS^-}^2(t)}.
\end{equation} 

\begin{figure*}
\includegraphics{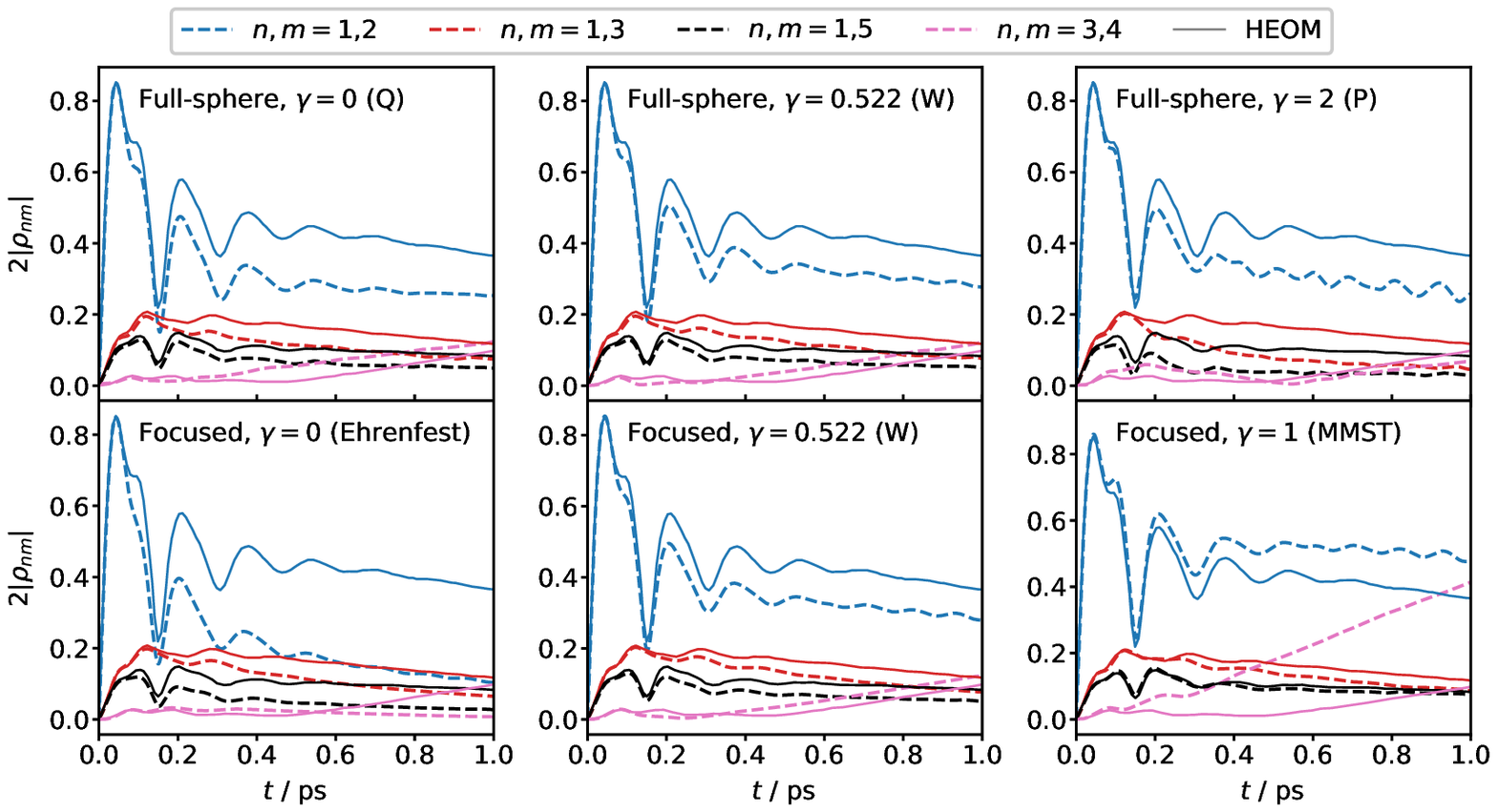}
\caption{Bipartite entanglement for the FMO model at $T=77$\,K with $\tau_\text{c}=100$\,fs, starting from state 1. Solid lines show numerically exact HEOM results.\cite{sarovar2010nature}}\label{fig:cohA1}
\end{figure*}

In Fig.~\ref{fig:cohA1} we show the concurrences that are largest in magnitude for an FMO model with $\tau_\text{c}=100$ fs. As before, the middle panels are the most accurate, while both Ehrenfest and focused MMST deviate significantly from the benchmark for at least one of the concurrences. All of the methods begin to disagree with the benchmark after about 0.2\,ps, but the error for W is smaller than what has previously been reported with PBME.\cite{kelly2011fmo}
It is noteworthy how the qualitative shapes of all lines can be predicted by our quasiclassical method, which cannot be done with Redfield theory (since that requires $\lambda$ to be much smaller than the electronic couplings).\cite{sarovar2010nature} In the Supplementary Material we also show the long-time limit of Fig.~\ref{fig:cohA1}, the equivalent calculations starting from state 6, as well as the higher temperature case, which illustrate the same trends as have already been pointed out.

All together, the symmetric W approach that we propose in this paper is seen to be a promising method that is both simple to compute and resolves several drawbacks of traditional MMST methods.

\section{Conclusions}
In this paper we have generalized the spin mapping of a two-level system in paper I\cite{runeson2019} to $N$-level systems, which is a problem that had not been satisfactorily solved since is was first posed by the seminal works of Meyer and Miller in 1979.\cite{meyer1979spin} 
The general idea is to make the classical phase space inherit the $SU(N)$-symmetry properties of the quantum system. Of particular significance is the Casimir invariant, which plays the role of a generalized (squared) spin magnitude. This quantity is independent of basis representation and controls the overall strength of the nuclear forces.%

In contrast with previous spin mapping attempts,\cite{meyer1979spin,Cotton2015,liu2016} we have shown how the dynamics can be generated by a quadratic Hamiltonian of the same form as in the standard harmonic-oscillator mapping, but with a new formula for the zero-point energy parameter $\gamma$. Originally $\gamma$ was included as a Langer correction, then justified through the commutation relations of a set of harmonic-oscillator operators. Now we recommend that this term is changed to become dependent on $N$, with values close to what was previously found optimal when it was treated as a free parameter.\cite{Mueller1999pyrazine}

One can therefore say that the generalized spin mapping is a more natural derivation of the MMST Hamiltonian than the original harmonic-oscillator mapping. We have shown that the spin mapping solves the problem of leakage from the physical space, so that there is no need for additional projectors. The present theory also does not assume any particular form of the Hamiltonian other than that the subsystem belongs to the symmetry group $SU(N)$, while MMST-based approaches often depend on how $\hat{V}(x)$ is split in a state-dependent and a state-independent part. 

We have demonstrated that the resulting method can predict population dynamics in benchmark systems to similar accuracy as other state-of-the-art mapping approaches such as SQC and traceless MMST. In the future we expect that the accuracy can be extended to longer times by combining the dynamics with a generalized quantum master equation, as has been successfully done for other mapping approaches.\cite{Kelly2016master,Reichman2016spinboson,Reichman2017spinboson,geva2019gqme} Another natural extension would be to develop an FBTS or PLDM method based on spin mapping. Finally, we also believe that the spin mapping will be relevant in the search for a nonadiabatic extension to ring-polymer molecular dynamics.\cite{RPMDcorrelation,mapping,richardson2017vibronic,Ananth2013MVRPMD,Chowdhury2017coherent}

\section*{Supplementary Material}
Supplementary Material is available with additional results for the FMO model.

\begin{acknowledgments}
The authors would like to thank Maximilian Saller for valuable advice on the FMO model. We also thank Gabriella Wallentin for testing various other model systems, as well as Jonathan Mannouch and Aaron Kelly for fruitful discussions.
J.E.R.\ is supported by the Hans H. G\"{u}nthard scholarship, %
and both authors
acknowledge support from the Swiss National Science Foundation through the NCCR MUST (Molecular Ultrafast Science and Technology) Network.
\end{acknowledgments}

\appendix

\section{The Casimir invariant}\label{sec:app_casimir}
Here we give a proof for the formula for the Casimir invariant that is known from many textbooks (for example p.~500 in Ref.~\onlinecite{peskin1995QFT}).
The generators of the Lie algebra $\mathfrak{su}(N)$ have commutation relations of the form
\begin{equation}
    [\hat{S}_i,\hat{S}_j] = \ii \sum_k f_{ijk} \hat{S}_k,
\end{equation}
where $f_{ijk}$ is totally antisymmetric and contains the \emph{structure constants} of $\mathfrak{su}(N)$. 
Now define the quadratic Casimir operator $\hat{C}_2\equiv \sum_{i=1}^{N^2-1} \hat{S}_i^2$. It is easy to show that it commutes with all generators:
\begin{multline}
 [\sum_i\hat{S}_i^2,\hat{S}_j] =\sum_i( \hat{S}_i[\hat{S}_i,\hat{S}_j] + [\hat{S}_i,\hat{S}_j]\hat{S}_i) \\ 
 = \ii \sum_{ik} f_{ijk}(\hat{S}_i\hat{S}_k + \hat{S}_k \hat{S}_i)=0,
\end{multline}
since $f_{ijk}=-f_{kji}$. Thus it must be proportional to the identity,
\begin{equation}
\hat{C}_2= C_2 \hat{\mathcal{I}},
\end{equation}
and the proportionality constant is easily found as
\begin{equation}\label{eq:CN}
C_2 = \frac{\sum_i \Tr[\hat{S}_i^2]}{\Tr[\hat{\mathcal{I}}]} = \frac{\sum_{i=1}^{N^2-1} \frac{1}{2}\delta_{ii}}{N} = \frac{N^2-1}{2N}.
\end{equation}

\section{Generalized spin operators in $N$-level problems}\label{sec:app_basis}
There always exists a set of traceless Hermitian $(N\times N)$-matrices that fulfils the properties in Sec.~\ref{sec:spinmatrices}. One common construction that is a direct generalization of the Pauli matrices consists of:\cite{hioe1981genGMmat}
\begin{subequations}
\begin{itemize}
    \item the $N(N-1)/2$ symmetric matrices
    \begin{equation}
        \hat{S}^+_{mn} = \thalf(|m\rangle\langle n| + |n\rangle\langle m|), \quad 1\leq m < n \leq N ,
    \end{equation}
    \item the $N(N-1)/2$ asymmetric matrices
    \begin{equation}
        \hat{S}^-_{mn} = -\tfrac{\ii}{2}(|m\rangle\langle n| - |n\rangle\langle m|), \quad 1\leq m < n \leq N ,
    \end{equation}
    \item the $N-1$ diagonal matrices
    \begin{equation}
        \hat{S}_n = \sqrt{\frac{1}{2n(n-1)}}\left(\sum_{k=1}^{n-1}|k\rangle\langle k|+(1-n)|n\rangle\langle n|\right), \quad 2\leq n \leq N .
    \end{equation}
\end{itemize}
\end{subequations}
Other bases can be constructed from linear combinations of these.
However, as previously mentioned, it is not actually necessary to choose a particular basis in order to obtain the results of this paper.

\section{Formulation in spin-1 matrices}\label{sec:app_spin1}
It is well known that an $N$-level system can also be described in terms of operators of a spin $S=\thalf(N-1)$ system.\cite{fano1983}
Here we show the specific example of how a three-level system can be related to a spin-1 particle, in a slightly different way than Meyer and Miller in Ref.~\onlinecite{meyer1979spin}. 

One way to represent a spin-1 is via symmetrized product states of two spin-$\thalf$ particles. A natural basis of this (triplet) space is $\{ \ket{\uparrow \uparrow}, \tfrac{1}{\sqrt{2}}(\ket{\uparrow\downarrow}+\ket{\downarrow\uparrow}),\ket{\downarrow\downarrow} \}$. In this basis, the total spin projection 
along each of the coordinate axes are
\begin{subequations}
\begin{align}
    \hat{S}_1 &= \hat{S}_x = \frac{1}{2\sqrt{2}} \begin{pmatrix} 0 & 1 & 0 \\ 1 & 0 & 1 \\ 0 & 1 & 0 \end{pmatrix},  \\
    \hat{S}_2 &= \hat{S}_y = \frac{1}{2\sqrt{2}} \begin{pmatrix} 0 & -\ii & 0 \\ \ii & 0 & -\ii \\ 0 & \ii & 0 \end{pmatrix}, \\
    \hat{S}_3 &= \hat{S}_z = \frac{1}{2} \begin{pmatrix} 1 & 0 & 0 \\ 0 & 0 & 0 \\ 0 & 0 & -1 \end{pmatrix},
\end{align}
\end{subequations}
which we can take as the first three basis matrices. Their phase-space functions are analogous to $p$-orbitals. As the remaining five basis matrices, we take the following $d$-orbital analogues:
\begin{subequations}
\begin{align}
    \hat{S}_4 &= 2(\hat{S}_x\hat{S}_z+\hat{S}_z\hat{S}_x) =\frac{1}{2} \begin{pmatrix} 0 & 0 & -\ii \\ 0 & 0 & 0 \\ \ii & 0 & 0 \end{pmatrix}, \\
    \hat{S}_5 &= 2(\hat{S}_x\hat{S}_y+\hat{S}_y\hat{S}_x) =\frac{1}{2\sqrt{2}} \begin{pmatrix} 0 & -\ii & 0 \\ \ii & 0 & \ii \\ 0 & -\ii & 0 \end{pmatrix}, \\
    \hat{S}_6 &= 2(\hat{S}_y\hat{S}_z+\hat{S}_z\hat{S}_y)  =\frac{1}{2\sqrt{2}} \begin{pmatrix} 0 & 1 & 0 \\ 1 & 0 & 1 \\ 0 & 1 & 0 \end{pmatrix},  \\
    \hat{S}_7 &= 2(\hat{S}_x^2-\hat{S}_y^2) = \frac{1}{2} \begin{pmatrix} 0 & 0 & 1 \\ 0 & 0 & 0 \\ 1 & 0 & 0 \end{pmatrix},\\
    \hat{S}_8 &= \frac{2}{\sqrt{3}}(2\hat{S}_z^2-\hat{S}_x^2-\hat{S}_y^2) = \frac{1}{2\sqrt{3}} \begin{pmatrix} 1 & 0 & 0 \\ 0 & -2 & 0 \\ 0 & 0 & 1 \end{pmatrix},
\end{align}
\end{subequations}
which are clearly linear combinations of the Gell-Mann matrices in Eq.~\eqref{eq:GellMann}, but with a physical meaning in terms of spin-1.%

These matrices are subtly different from those of Meyer and Miller in Ref.~\onlinecite{meyer1979spin} which defined $\hat{S}_8$ as $\hat{S}_z^2$. This has a non-zero trace and therefore does not comply with our requirements.

\section{Generalized coherent states in spherical variables}\label{sec:app_coh}
One can construct the coherent states in many ways.\cite{perelomov1986book,nemoto2000} A simple way is to define the $N$-level coherent state $|\Omega\rangle_N$ iteratively through
\begin{equation}
\langle n|\Omega\rangle_N =
\begin{cases}
    \langle n|\Omega\rangle_{N-1} & 1\leq n<N-1\\
    \langle N-1|\Omega\rangle_{N-1}\cos\tfrac{\theta_N}{2} & n=N-1 \\
    \langle N-1|\Omega\rangle_{N-1}\eu{\ii\varphi_N}\sin\tfrac{\theta_N}{2} &n=N
\end{cases}
\end{equation}
starting from $\langle 1|\Omega\rangle_1 = 1$. This gives the $N=2$ coherent state
\begin{equation}
    |\Omega\rangle_2 = \begin{pmatrix} \cos\tfrac{\theta_1}{2} \\ \eu{\ii\varphi_1}\sin\tfrac{\theta_1}{2} \end{pmatrix}
\end{equation}
which coincides with Eq.~\eqref{eq:coherent_state2} up to a global phase, which of course does not affect the values of $\langle \Omega|\hat{S}_i|\Omega\rangle$. In the $N=3$ case we get
\begin{equation}
    |\Omega\rangle_3 = \begin{pmatrix} \cos\tfrac{\theta_1}{2} \\ \eu{\ii \varphi_1} \sin\tfrac{\theta_1}{2}\cos\tfrac{\theta_2}{2} \\ \eu{\ii(\varphi_1+\varphi_2)}\sin\tfrac{\theta_1}{2}\sin\tfrac{\theta_2}{2} \end{pmatrix},
\end{equation}
and so on. Note that the coherent states are always normalized such that $\langle \Omega|\Omega\rangle=1$. The $2N-2$ angular variables have the domains $0\leq \theta_n \leq \pi$ and $0\leq \varphi_n \leq 2\pi$. The differential phase-space volume element is\cite{tilma2004eulerangle}
\begin{equation}
    \rd\Omega = \frac{N!}{2(2\pi)^{N-1}} \prod_{1\leq n \leq N-1} K_n(\theta_n,\varphi_n)\rd\theta_n\rd\varphi_n,
\end{equation}
with
\begin{equation*}
    K_n(\theta_n,\varphi_n) = 
    \begin{cases} 
        \sin\theta_n & n=1 \\
        \left(\cos\tfrac{\theta_n}{2}\right)^{2n-1}\sin\tfrac{\theta_n}{2} & 1<n<N \\
        \cos\tfrac{\theta_n}{2}\left(\sin\tfrac{\theta_n}{2}\right)^{2N-3} & n=N-1 \text{ and } N > 2.
    \end{cases}
\end{equation*}
The coherent states allow for a resolution of unity:\cite{klauder1985book}
\begin{equation}
    \hat{\mathcal{I}}=\int \rd\Omega\, |\Omega\rangle\langle \Omega|. 
\end{equation}
By taking the trace of each side, it is clear that $\int\rd\Omega = N$.

As an example for $N=3$, the orthogonal functions for the Gell-Mann matrices in Eq.~\eqref{eq:GellMann} are
\begin{subequations}
\begin{align}
    \langle \Omega|\hat{S}_1|\Omega\rangle &= \thalf \sin\theta_1 \cos\tfrac{\theta_2}{2} \cos\varphi_1 \\
    \langle \Omega|\hat{S}_2|\Omega\rangle &= \thalf \sin\theta_1 \cos\tfrac{\theta_2}{2} \sin\varphi_1 \\
    \langle \Omega|\hat{S}_3|\Omega\rangle &= \tfrac{1}{8}(1+3\cos\theta_1-\cos\theta_2+\cos\theta_1\cos\theta_2) \\
    \langle \Omega|\hat{S}_4|\Omega\rangle &= \thalf \sin\theta_1 \sin\tfrac{\theta_2}{2} \cos(\varphi_1+\varphi_2) \\
    \langle \Omega|\hat{S}_5|\Omega\rangle &= \thalf \sin\theta_1 \sin\tfrac{\theta_2}{2} \sin(\varphi_1+\varphi_2) \\
    \langle \Omega|\hat{S}_6|\Omega\rangle &= \thalf \sin^2\tfrac{\theta_1}{2}\sin\theta_2\cos\varphi_2 \\
    \langle \Omega|\hat{S}_7|\Omega\rangle &= \thalf \sin^2\tfrac{\theta_1}{2}\sin\theta_2\sin\varphi_2 \\
    \langle \Omega|\hat{S}_8|\Omega\rangle &= \tfrac{1}{8\sqrt{3}}(1+3\cos\theta_1+3\cos\theta_2-3\cos\theta_1\cos\theta_2),
\end{align}
\end{subequations}
and it is easy to check that they fulfil the orthogonality relation in Eq.~\eqref{eq:orthogonalityN}. A more detailed phase-space treatment of the $N=3$ case can be found in Ref.~\onlinecite{luis2008}.

In these variables the W-representation of the single-state projectors are
\begin{subequations}
\begin{align}
    [|1\rangle\langle 1|]_\text{W}&=\frac{2}{3}+\cos\theta_1 \\
    [|2\rangle\langle 2|]_\text{W}&=\frac{1}{6}(1 - 3\cos\theta_1 + 3\cos\theta_2 - 3\cos\theta_1\cos\theta_2) \\
    [|3\rangle\langle 3|]_\text{W}&=\frac{1}{6}(1 - 3\cos\theta_1 - 3\cos\theta_2 + 3\cos\theta_1\cos\theta_2).
\end{align}
\end{subequations}
Note that their sum is one for all angles, meaning that the total population is identically one. To find when the system is entirely in state $n$, we solve the system of equations $[|n\rangle\langle n|]_\text{W}=1$ and $[|k\rangle\langle k|]_\text{W}=0$ for $k\neq n$. The solutions define circles with fixed $\theta_k$ for $k\leq n$ (and $k\leq N-1$), see Fig.~\ref{fig:spheres}.

Lastly we point out that even though we visualize the spherical coordinates using multiple ``spins'', the phase space constructed from $SU(N)$-symmetric coherent states is different from when mapping multiple spins independently to a classical phase space (as in Ref.~\onlinecite{fay2019faraday}).

\section{Auxiliary formulas}\label{sec:app_aux}
To prove the preservation of the Casimir invariant in Eq.~\eqref{eq:SWsum}, we need some further properties of the coherent states. It is known from the theory of harmonic functions that $|\langle\Omega| \Omega'\rangle|^2$ can be expanded as\cite{brif1999phase}
\begin{equation}\label{eq:Yexpansion}
    |\langle\Omega| \Omega'\rangle|^2 = \sum_{\nu=0}^{N^2-1} \tau_\nu Y_\nu^*(\Omega)Y_\nu(\Omega'),
\end{equation}
where $\tau_\nu$ are constants and $Y_\nu(\Omega)=\left(\frac{2}{\tau_\nu}\right)^{1/2} \langle\Omega|\hat{S}_\nu|\Omega\rangle$ are generalized spherical harmonics that fulfil $\int \rd\Omega\,Y_\nu(\Omega) Y_{\nu'}(\Omega)=\delta_{\nu\nu'}$. The index $\nu$ runs from $0$ to $N^2-1$, where $\hat{S}_0=\frac{1}{\sqrt{2N}}\hat{\mathcal{I}}$. One can show that $\tau_\nu$ is invariant to transformations within an irreducible subspace,\cite{brif1999phase} which for our purposes means that $
\{\tau_i\}_{i=1}^{N^2-1}$ are all equal (to, say, $\tau_1$), and only $\tau_0$ is different.\footnote{It is clear from $\bra{\Omega}\hat{\mathcal{I}}\ket{\Omega}=1$ that $Y_0(\Omega)=1/\sqrt{N\tau_0}$. The normalization $\int\rd\Omega\,|Y_0(\Omega)|^2=1$ then directly gives $\tau_0=1$ and $Y_0(\Omega)=1/\sqrt{N}$. }

Let us insert $Y_\nu(\Omega)$ into Eq.~\eqref{eq:Yexpansion} and set $\Omega'=\Omega$:
\begin{equation}
    1 = \left(\frac{1}{\sqrt{N}}\right)^2 + \sum_{i=1}^{N^2-1} 2 \langle \Omega|\hat{S}_i|\Omega\rangle^2,
\end{equation}
so that
\begin{equation}
    \sum_{i=1}^{N^2-1}  \langle \Omega|\hat{S}_i|\Omega\rangle^2 = \frac{N-1}{2N},
\end{equation}
which is used in Eq.~\eqref{eq:SWsum}. Further, the normalization of $Y_i(\Omega)$ gives 
\begin{equation}\label{eq:orthogonality_aux}
    \int \rd\Omega\,  \langle \Omega|\hat{S}_i|\Omega\rangle \langle \Omega|\hat{S}_j|\Omega\rangle = \thalf \tau_1\delta_{ij}.
\end{equation}
Summation over all $i$ and $j$ leads to
\begin{equation}
    N^2-1 = \frac{2}{\tau_1} \int \rd\Omega \sum_{i=1}^{N^2-1} \langle \Omega|\hat{S}_i|\Omega\rangle^2 = \frac{2}{\tau_1}N\frac{N-1}{2N},
\end{equation}
giving $\tau_1=1/(N+1)$. Insertion into Eq.~\eqref{eq:orthogonality_aux} finally gives the orthogonality relation in Eq.~\eqref{eq:orthogonalityN}.

\bibliography{references,johanrefs}%

\end{document}


\title{Generalized spin mapping for quantum-classical dynamics: Supplementary Material} %

\author{Johan E. Runeson}
\email{johan.runeson@phys.chem.ethz.ch}
\author{Jeremy O. Richardson}%
\email{jeremy.richardson@phys.chem.ethz.ch}
\affiliation{Laboratory of Physical Chemistry, ETH Z\"{u}rich, 8093 Z\"{u}rich, Switzerland}

\date{\today}%

\begin{abstract}
    Here we show additional results for the FMO model using the spin-mapping methods explained in the main paper. For the population dynamics, these include the case of starting from state 6 (instead of state 1), and the case of a slower bath. For the bipartite entanglement we show the long-time results for the system in the main paper, as well as a system at high temperature and the case of starting from state 6.
    Together these figures further strengthen the conclusion of the main paper that the W-representation gives the most reliable results, both when using full-sphere and focused initial conditions.
\end{abstract}

\maketitle

\begin{figure*}%
\includegraphics{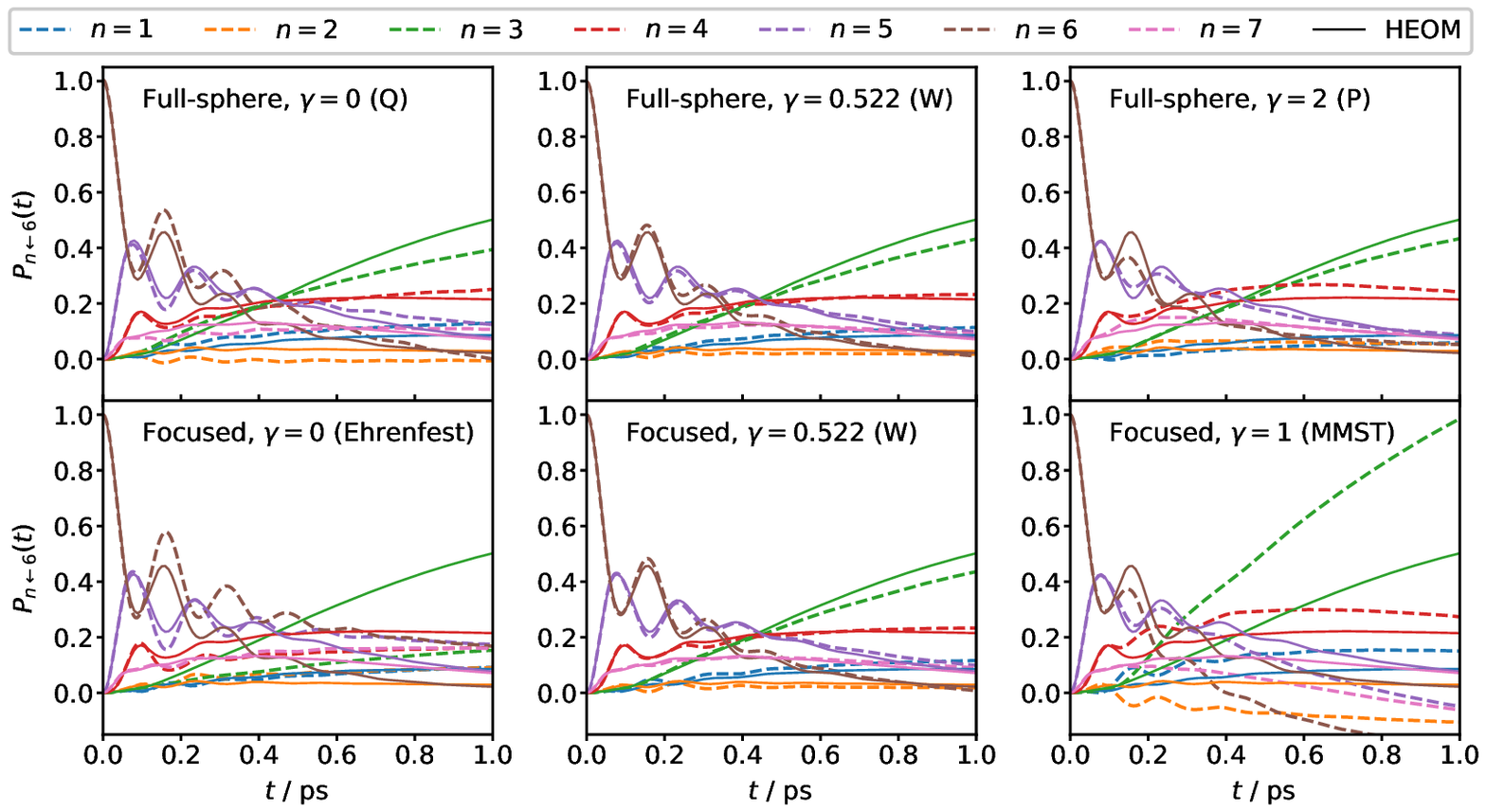}
\caption{Same as Fig.~2 in the paper ($T=\SI{77}{K}$, $\tau_\text{c}=\SI{50}{fs}$) but starting from state 6. Solid lines show numerically exact HEOM results.\cite{ishizaki2009FMO} The system is the same as in Fig.~11a of Ref.~\onlinecite{Cotton2019manystates}, Fig.~4 of Ref.~\onlinecite{cotton2016lightharvest}, and Figs.~2--3 of Ref.~\onlinecite{saller2020faraday}.}\label{fig:A6}
\end{figure*}

\begin{figure*}%
\includegraphics{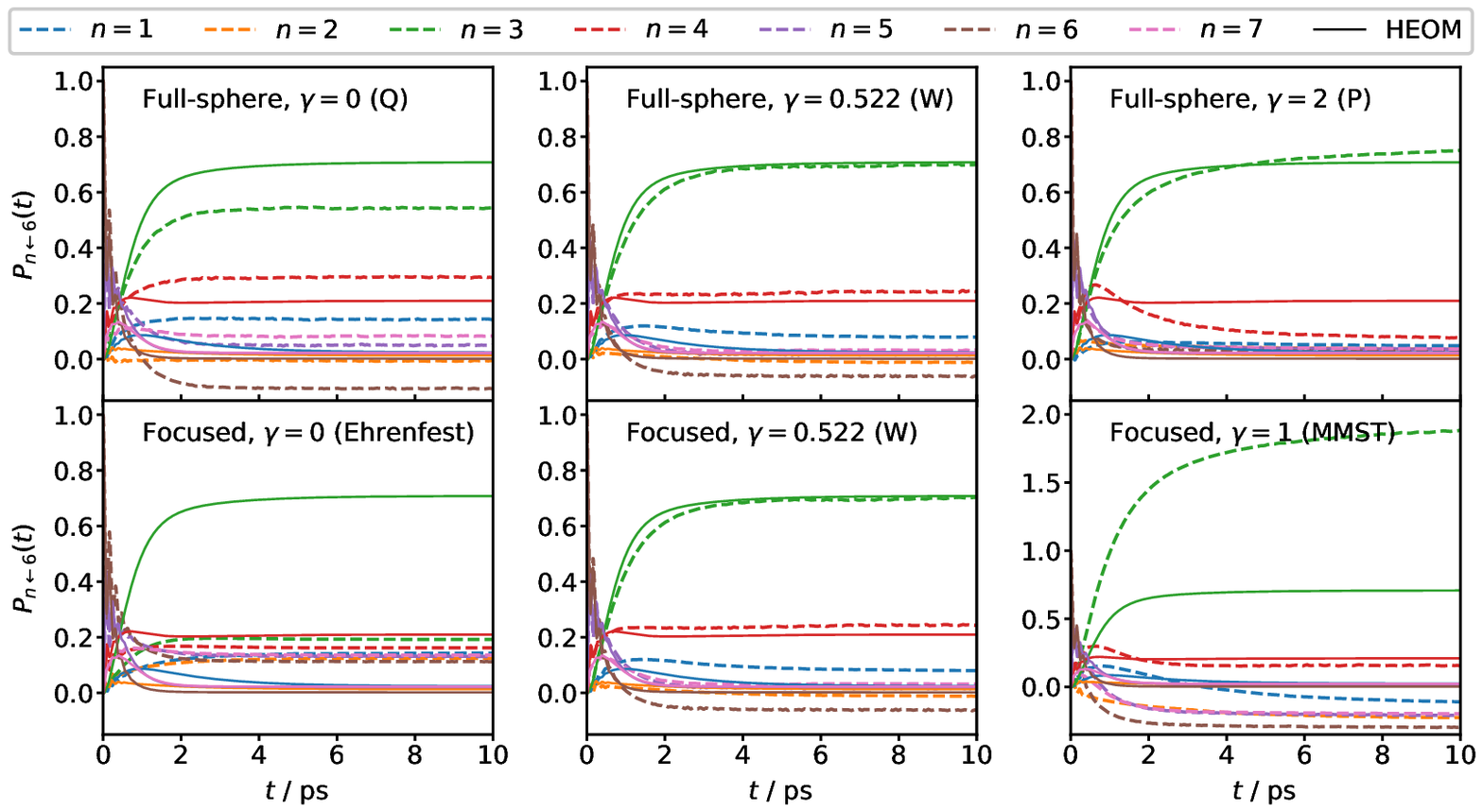}
\caption{Same as Fig.~3 in the paper ($T=\SI{77}{K}$, $\tau_\text{c}=\SI{50}{fs}$) but starting from state 6. Solid lines show numerically exact HEOM results.\cite{dattani2015fmo} The system is the same as in Fig.~6 of Ref.~\onlinecite{saller2020faraday}.}\label{fig:A6_long}
\end{figure*}

\begin{figure*}%
\includegraphics{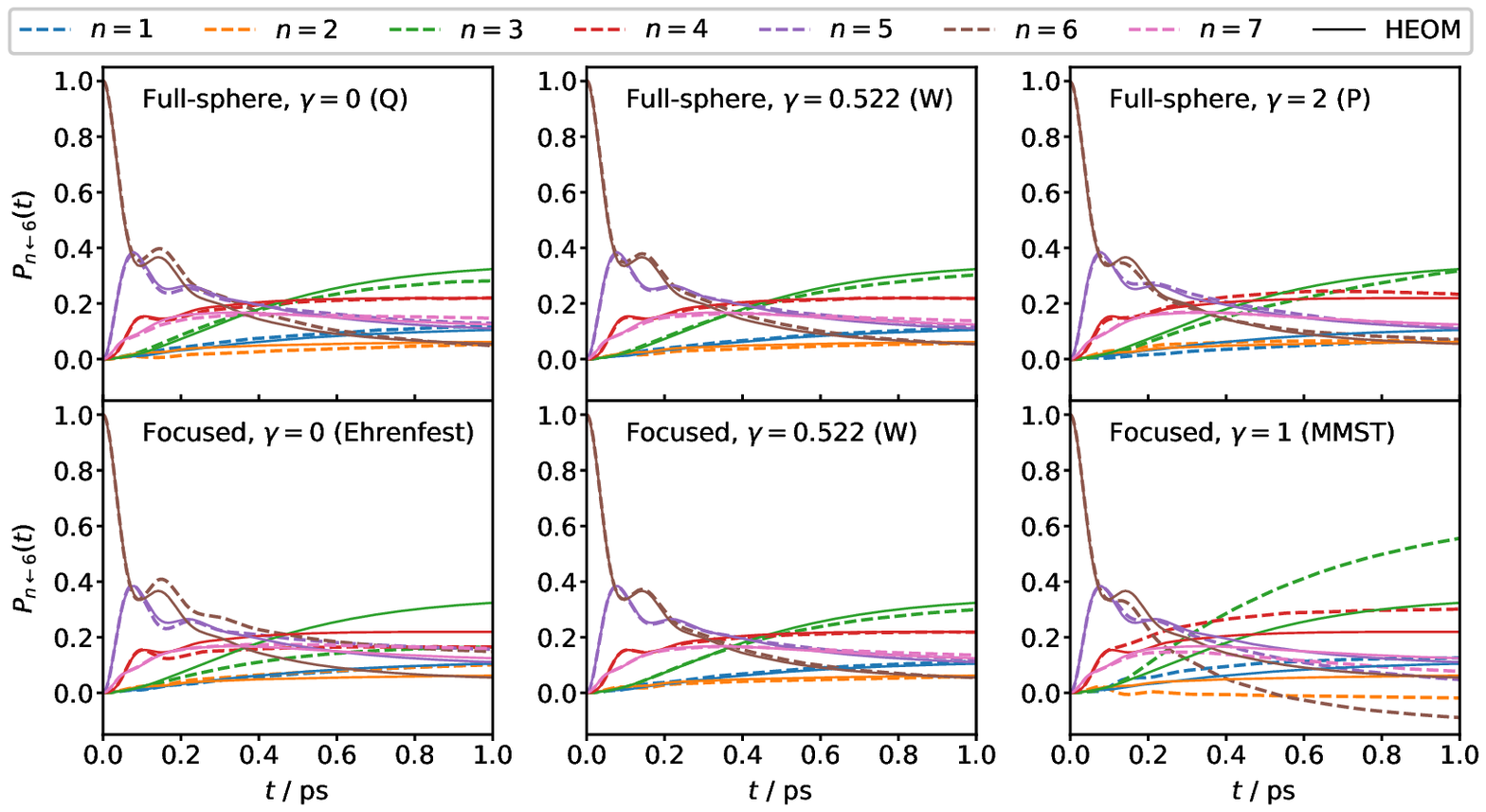}
\caption{Same as Fig.~4 in the paper ($T=\SI{300}{K}$, $\tau_\text{c}=\SI{50}{fs}$) but starting from state 6. Solid lines show numerically exact HEOM results.\cite{ishizaki2009FMO} The system is the same as in Fig.~11c of Ref.~\onlinecite{Cotton2019manystates}, and Fig.~4 of Ref.~\onlinecite{saller2020faraday}.}\label{fig:B6}
\end{figure*}

\begin{figure*}%
\includegraphics{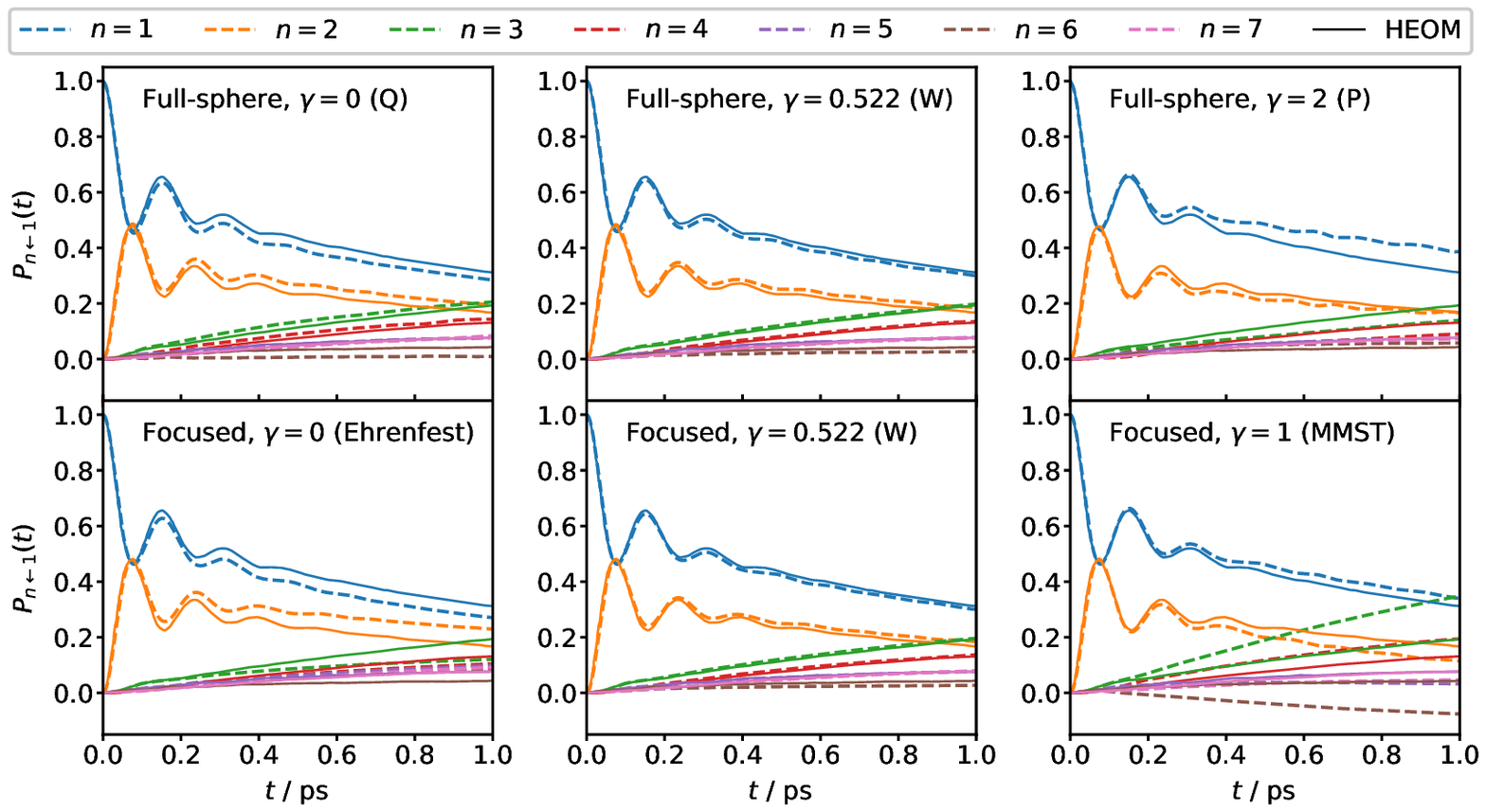}
\caption{Results at $T=300$\,K with $\tau_\text{c}=166$ fs (slower bath), starting from state 1. Solid lines show numerically exact HEOM results.\cite{ishizaki2009FMO} The system is the same as in Fig.~5 of Ref.~\onlinecite{saller2020faraday}. }\label{fig:C1_short}
\end{figure*}

\begin{figure*}
\includegraphics{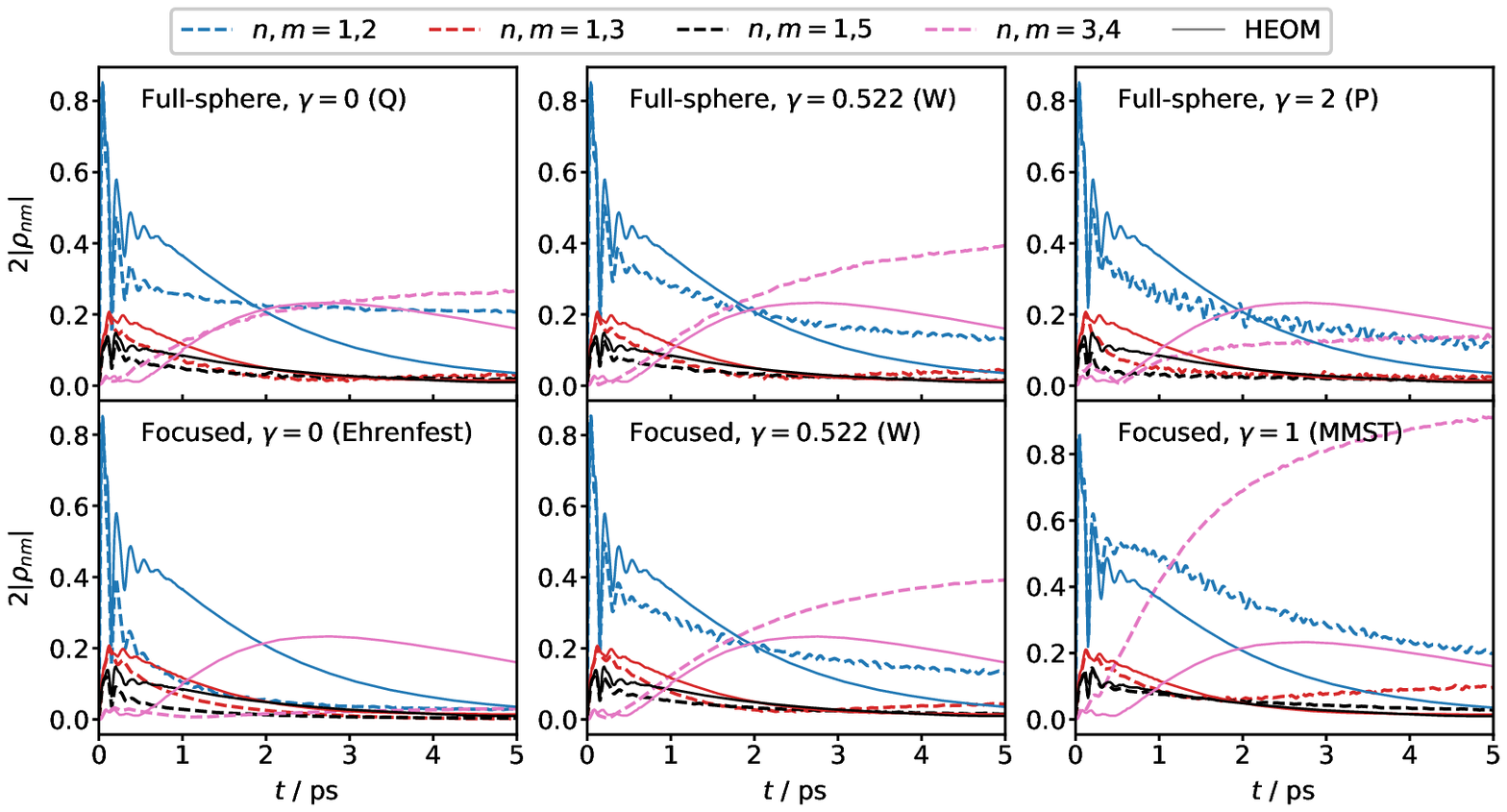}
\caption{Same as Fig.~5 in the paper ($T=77$\,K, $\tau_\text{c}=100$\,fs) but for longer times. Solid lines show numerically exact HEOM results.\cite{sarovar2010nature}}\label{fig:cohA1long}
\end{figure*}

\begin{figure*}
\includegraphics{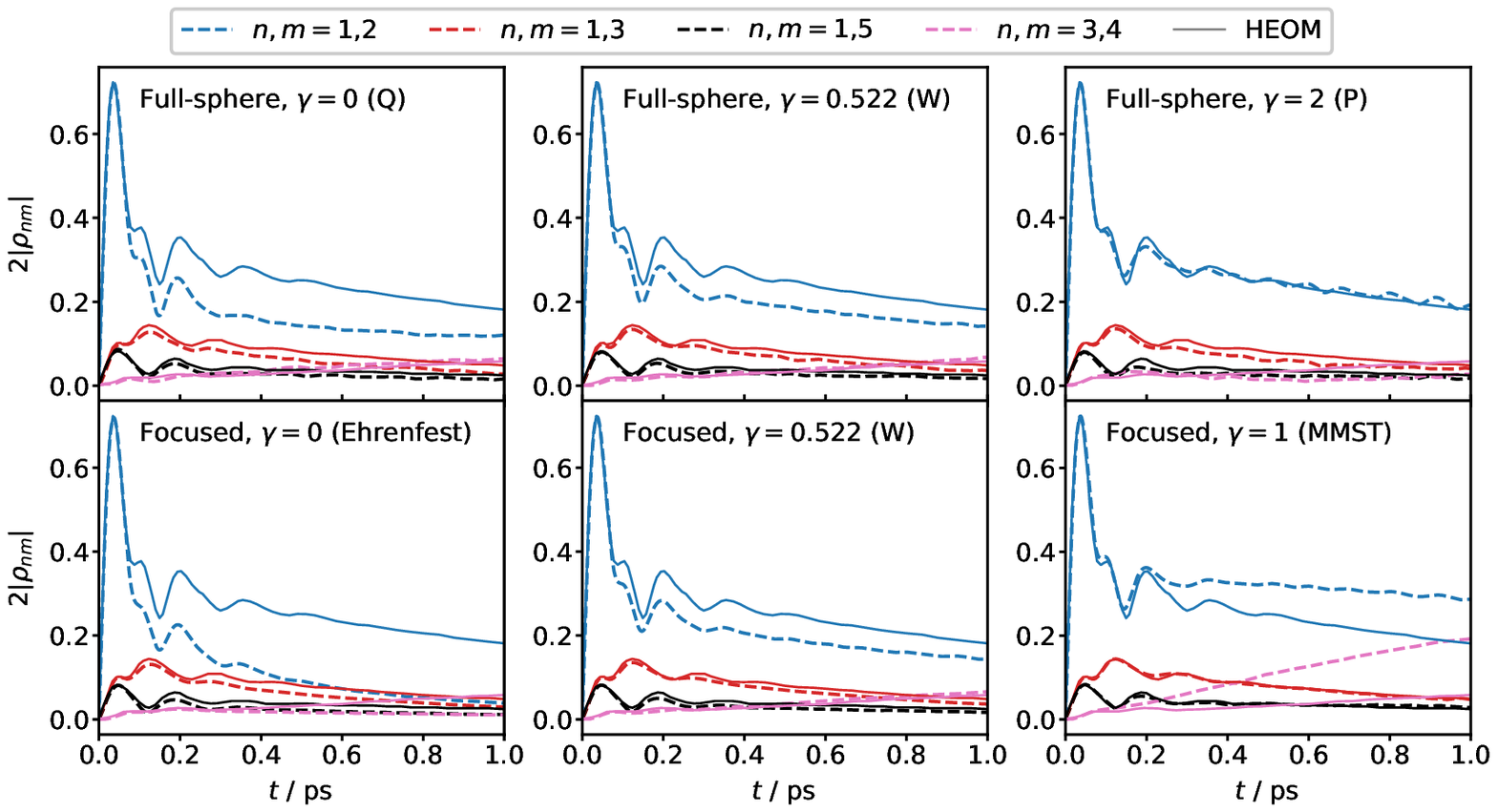}
\caption{Bipartite entanglement for the FMO model at $T=300$\,K with $\tau_\text{c}=100$ fs, starting from state 1. Solid lines show numerically exact HEOM results.\cite{sarovar2010nature}}\label{fig:cohB1}
\end{figure*}

\begin{figure*}
\includegraphics{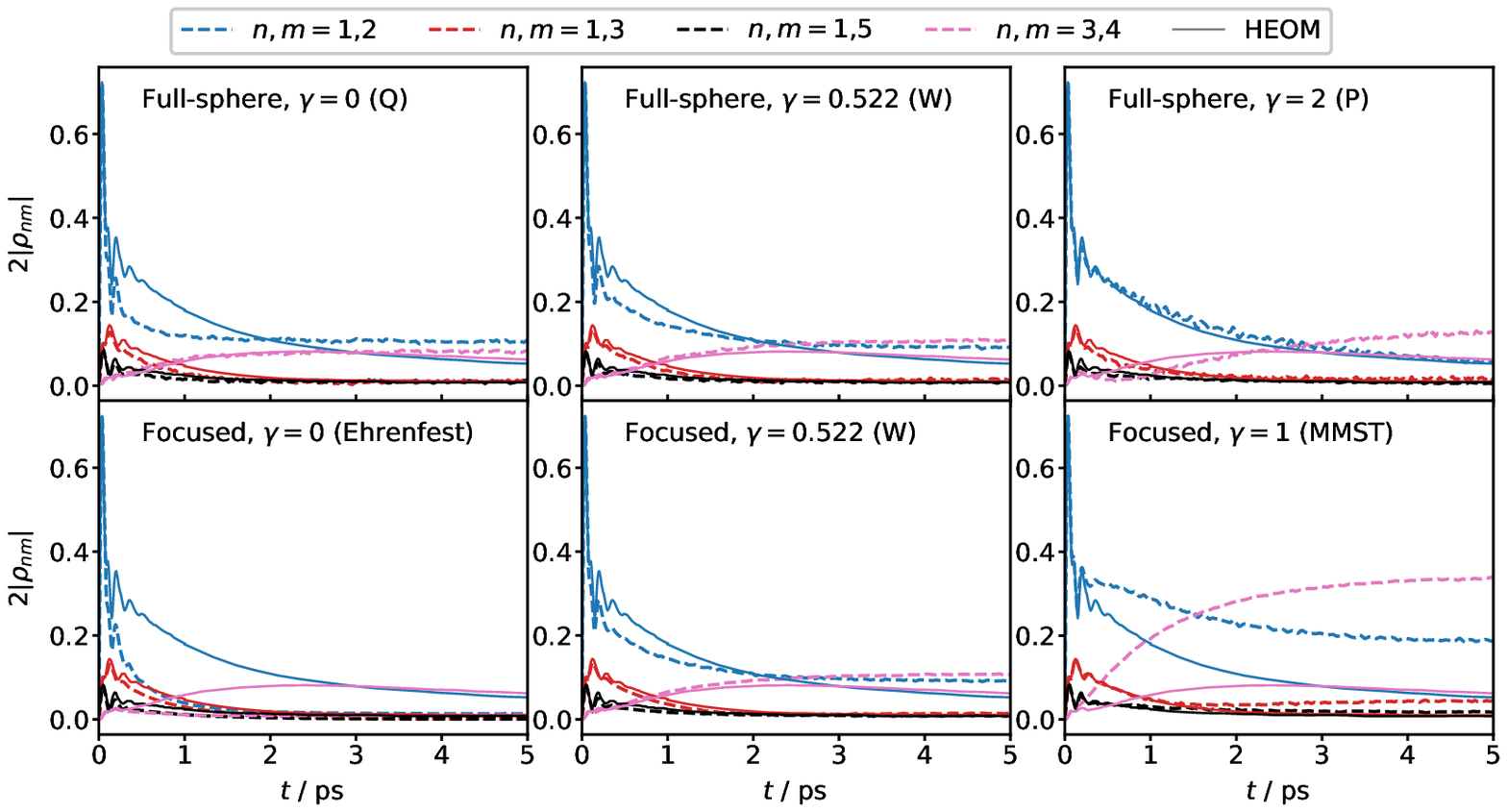}
\caption{Same as Fig.~\ref{fig:cohB1} but for longer times.  Solid lines show numerically exact HEOM results.\cite{sarovar2010nature}}\label{fig:cohB1long}
\end{figure*}

\begin{figure*}
\includegraphics{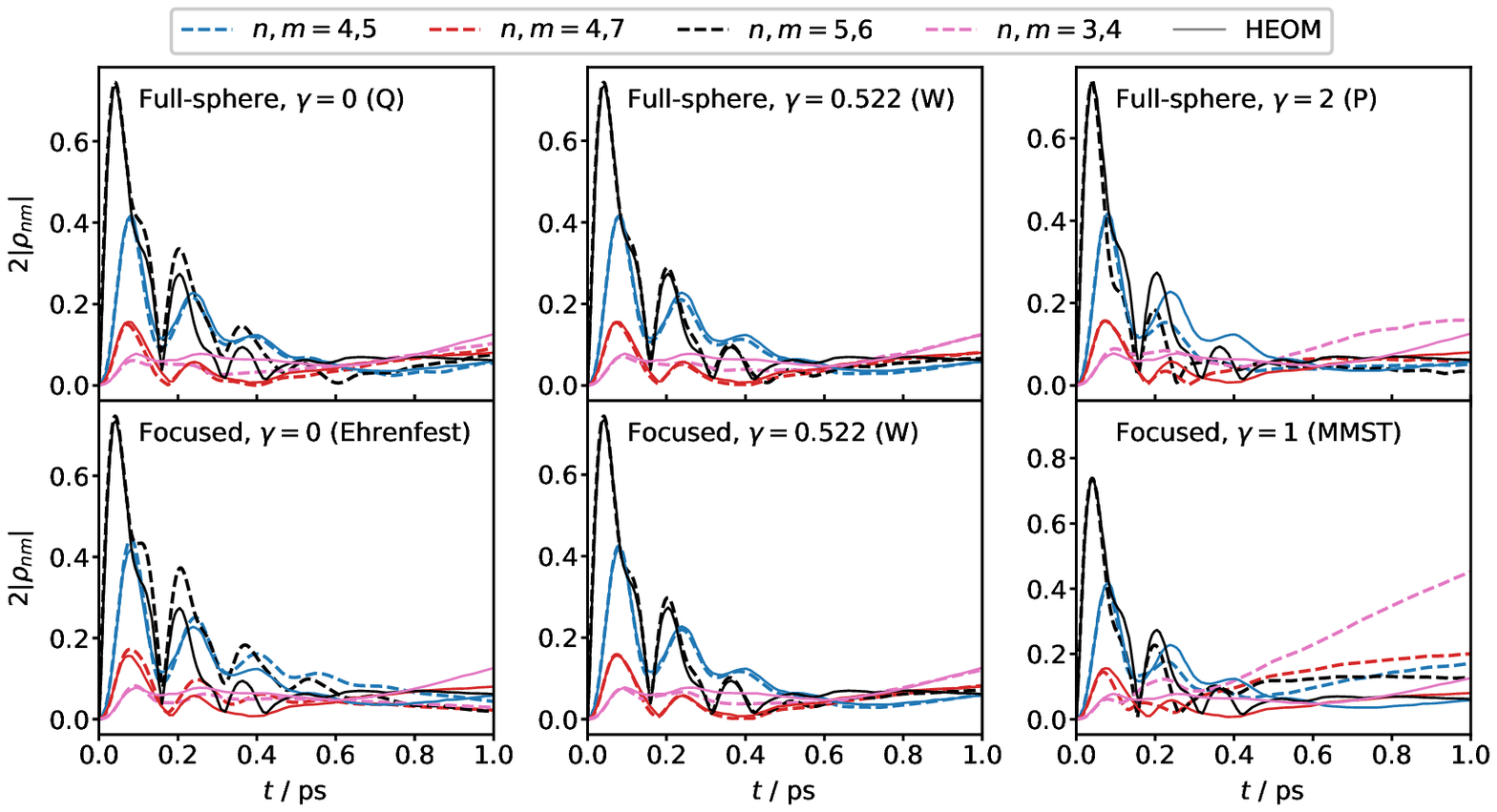}
\caption{Bipartite entanglement for the FMO model at $T=77$\,K with $\tau_\text{c}=100$ fs, starting from state 6. Solid lines show numerically exact HEOM results.\cite{sarovar2010nature}}\label{fig:cohA6}
\end{figure*}

\begin{figure*}
\includegraphics{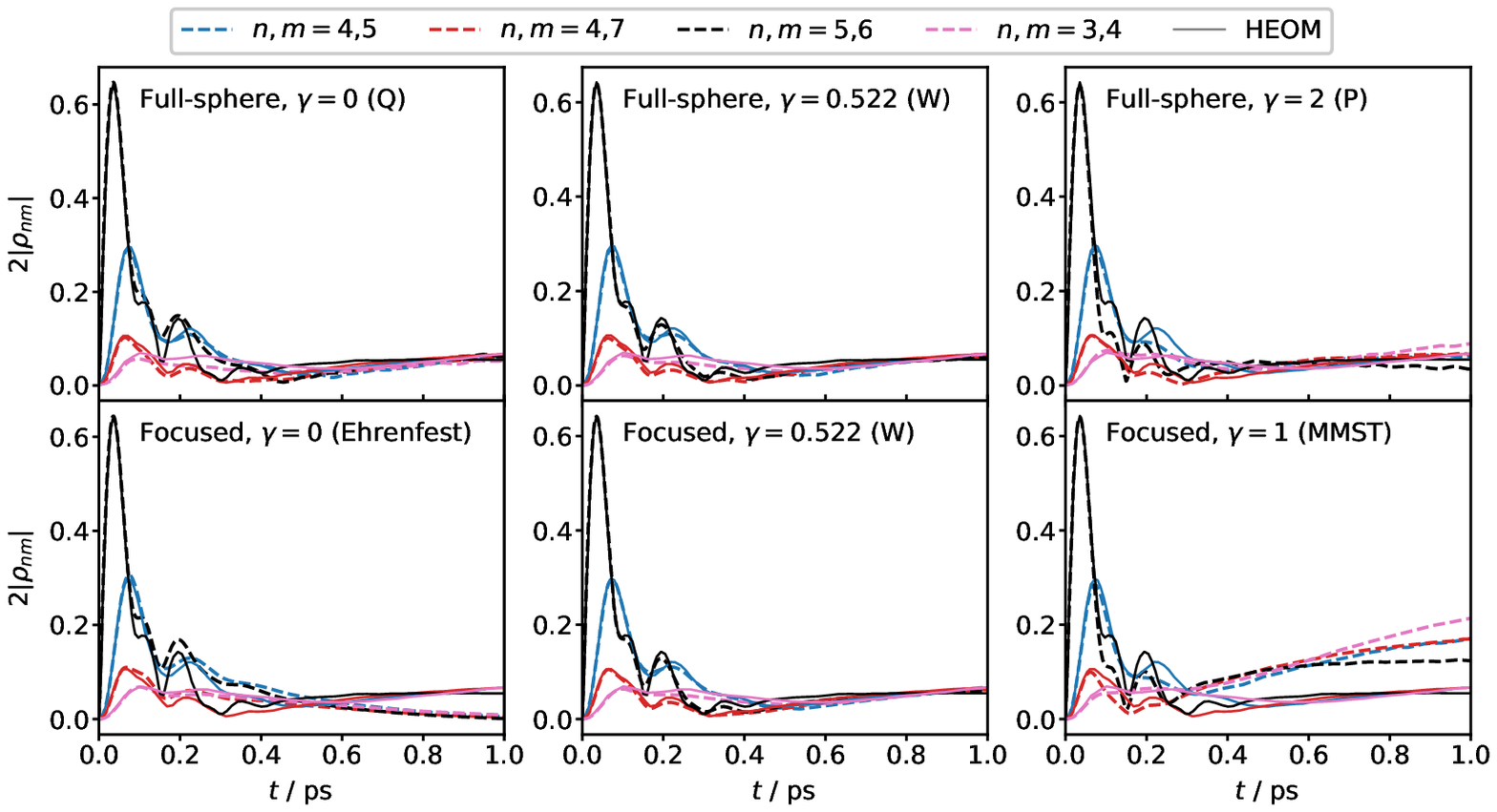}
\caption{Bipartite entanglement for the FMO model at $T=300$\,K with $\tau_\text{c}=100$ fs, starting from state 6. Solid lines show numerically exact HEOM results.\cite{sarovar2010nature}}\label{fig:cohB6}
\end{figure*}

\bibliography{references,johanrefs}